\documentclass[amsmath,amssymb,showpacs,showkeywords,twocolumn]{revtex4}
\usepackage[dvips]{graphicx,color}
\usepackage{times}
\usepackage{mathrsfs}
\usepackage{amsmath}
\usepackage{graphicx}
\usepackage{epsfig}
\usepackage{dcolumn}
\usepackage{bm}
\begin{document}
\title{Effect of Rashba spin-orbit coupling, magnetization and mixing of gap 
parameter on tunnelling conductance in F$|$NCSC junction of an F$|$S$|$F spin 
valve}
\author{Saumen Acharjee\footnote{saumenacharjee@gmail.com} and Umananda Dev 
Goswami\footnote{umananda2@gmail.com}}
\affiliation{Department of Physics, Dibrugarh University, Dibrugarh 786 004, 
Assam, India,}

\begin{abstract}
In this paper, we study the quantum transport at the 
Ferromagnet$|$Noncentrosymmetric Superconductor (F$|$NCSC) interface of an 
F$|$S$|$F spin valve. In this context, we investigate the tunneling 
conductance and its dependence on Rashba Spin-Orbit Coupling (RSOC) 
considering different barrier strength and a significant Fermi Wave-vector 
Mismatch (FWM) at the ferromagnetic and superconducting regions.  The study is 
carried out for different magnetization orientations and its strength. We 
developed Bogoliubov de Gennes (BdG) Hamiltonian introducing RSOC and exchange 
interaction for such an hybrid structure. To study charge conductance we use 
an extended Blonder -  Tinkham - Klapwijk (BTK) approach along with scattering 
matrix formalism to calculate the scattering coefficients. Our results 
strongly suggest that the tunneling conductance is strongly dependent on RSOC, 
magnetization strength, its orientation and the FWM. The work has also been 
done for different singlet-triplet mixing of the gap parameter. We have 
observed that with the rise of singlet-triplet mixing ratio the conductance 
decreases. It is also observed that a transparent barrier with 
moderate RSOC and having moderate magnetization strength with arbitrary 
orientation is highly suitable for maximum conductance. 
\end{abstract}

\pacs{67.30.hj, 85.75.-d, 74.90.+n}

\maketitle

\section{Introduction}
During the last few decades many heavy fermion compounds have been 
discovered which shows unconventional superconductivity \cite{saxena, aoki, 
pfleiderer, huy, buzdin, zutic, bauer1, bauer2, bauer3, motoyama, kawasaki, 
akazawa, anand1, smidman, anand2, yuan, togano, badica, matthias, singh, 
pecharsky, hillier, bonalde, yogi, ali, xu, flouquet, nandi}. With the 
discovery of unconventional superconductivity, the study of quantum transport 
in Ferromagnet$|$Superconductor (F$|$S) hybrid structures and spin valves 
gain lots of attention during this period not only from the fundamental 
physics point of view but also from the applications purpose, as these 
heterostructure hold a great potential for the applications in the 
nano-technological spintronic devices. Tunneling spectroscopy at the F$|$S 
interface is found to be one of the most powerful tool to investigate the 
nature of the superconducting states. 
Over the years, the study reveals many information for both conventional as 
well as unconventional pairing symmetries in the superconductors. From the 
point of view of Cooper pairing, two symmetries have been found very important 
in superconducting state: the symmetry of inversion center and the time 
reversal. In absence of one of these, Cooper pairs will appear in an 
unconventional form. Although Noncentrosymmetric Superconductors (NCSC) are 
the candidate of prime concern over the last decade, the field received a 
significant boost only since the discovery of noncentrosymmetry in heavy 
fermion compound CePt$_3$Si \cite{bauer1,bauer2,bauer3}. Soon, many 
superconducting materials had been 
reported which lacks center of inversion \cite{motoyama, kawasaki, akazawa, 
anand1, smidman, anand2, yuan, togano,badica, matthias, singh, pecharsky, 
hillier, bonalde, yogi, ali, xu}. A few of them are LaPt$_3$Si, UIr, 
La(Rh,Pt,Pd,Ir)Si$_3$, Li$_2$(Pt,Pd)$_3$B, Re$_6$Zr, LaNiC$_2$, 
Cd$_2$Re$_2$O$_7$, PbTaSe$_2$, etc. Due to the lack of center of symmetry, 
crystal structure induces an Antisymmetric Spin-Orbit Coupling (ASOC) and as 
a result parity no longer remains conserved. Consequently, the Fermi surface 
splits and the superconducting ground state exhibit an admixture of spin 
singlet and spin triplet components, if the pairing gap is much smaller 
than the Spin-Orbit Coupling (SOC) strength.

With the rise of spintronics and the ability to use spin degree of freedom 
with much precision in recent times, the field of spintronics gains a lot of 
attention from the application point of view. SOC is 
considered to be the central point of the emergence of spintronics. Rashba 
Spin-Orbit Coupling (RSOC) is a special kind of SOC. Unlike 
Dirac coupling it arises in crystals due to the lack of 
inversion symmetry. Since, the inversion center is absent in NCSC, hence it 
possess ASOC. So, it becomes really important to understand the role of RSOC 
in quantum transport and consequent tunneling process.

The tunneling conductance and its dependence on the SOC 
\cite{gorkov,yip,samokhin,sergienko,frigeri,fujimoto1,fujimoto2,molenkamp,wu,
wu1} had been studied previously for different hybrid structures. Recent 
theoretical researches indicate that the ratio of spin singlet to the spin 
triplet pairing states in a superconductor is highly anticipated by the 
strength of the SOC. This theoretical predictions was later 
supported by experimental findings in \cite{yuan, togano, badica} in heavy 
fermion compound Li$_2$(Pt,Pd)$_3$B. The experimental results 
showed that the pairing changes from spin-triplet state to spin-singlet state 
if Pt is replaced by Pd. Many other heavy fermion NCSC's had been studied 
since then and it was found that the pairing states were highly anticipated on 
strong SOC. Although conventional s-wave pairing \cite{bauer4,
wakui,kawasaki,ribeiro,biswas} is dominant in most cases, however it was also 
observed that the compounds with low spin-orbit coupling show unconventional 
superconductivity \cite{hillier,kuroiwa,chen1,chen2}. Hence, in this regard it 
becomes necessary to investigate the role of SOC on the 
pairing of NCSC's.
  
It is known from the earlier works that the role of magnetization in F$|$S 
hybrid structures \cite{cheng, cheng1, zutic1, zutic2,linder,
iniotakis,kashiwaya,banerjee,acharjee, linder1,bozovic1,bozovic2,tanaka}
is too significant. The transport properties and tunneling conductance in 
F$|$S (singlet) junction \cite{linder1,bozovic1,bozovic2}, 
F$|$S (triplet) junctions \cite{cheng, cheng1,zutic1,zutic2,tanaka} 
and NCSC structures \cite{linder,iniotakis,kashiwaya} have been studied earlier.
However, the interplay of magnetization with RSOC and its effect on the 
tunneling conductance in F$|$NCSC hybrid structures is still unknown. Since 
NCSC consists of both triplet and singlet pairing states simultaneously, thus 
it is necessary to investigate the interplay of magnetization and the 
mixed pairing states (singlet and triplet) in NCSC's and their influence on 
conductance characteristics.

Motivated by the earlier works as mentioned above, in this work we investigate
the effect of RSOC, magnetization and singlet-triplet mixing of gap parameter 
on the tunneling conductance due to the effect of these parameters on transport 
of an electron at the F$|$NCSC junction of an F$|$S$|$F spin valve.
 
The paper is organized as follows. In the Section II, a theoretical framework 
of the proposed setup is developed. The results of our work is discussed in 
the Section III. Finally we conclude our work in the Section IV.

\section{Theory}
\subsection{Model and formalism}
\begin{figure}[hbt]
\centerline
\centerline{ 
\includegraphics[scale=0.27]{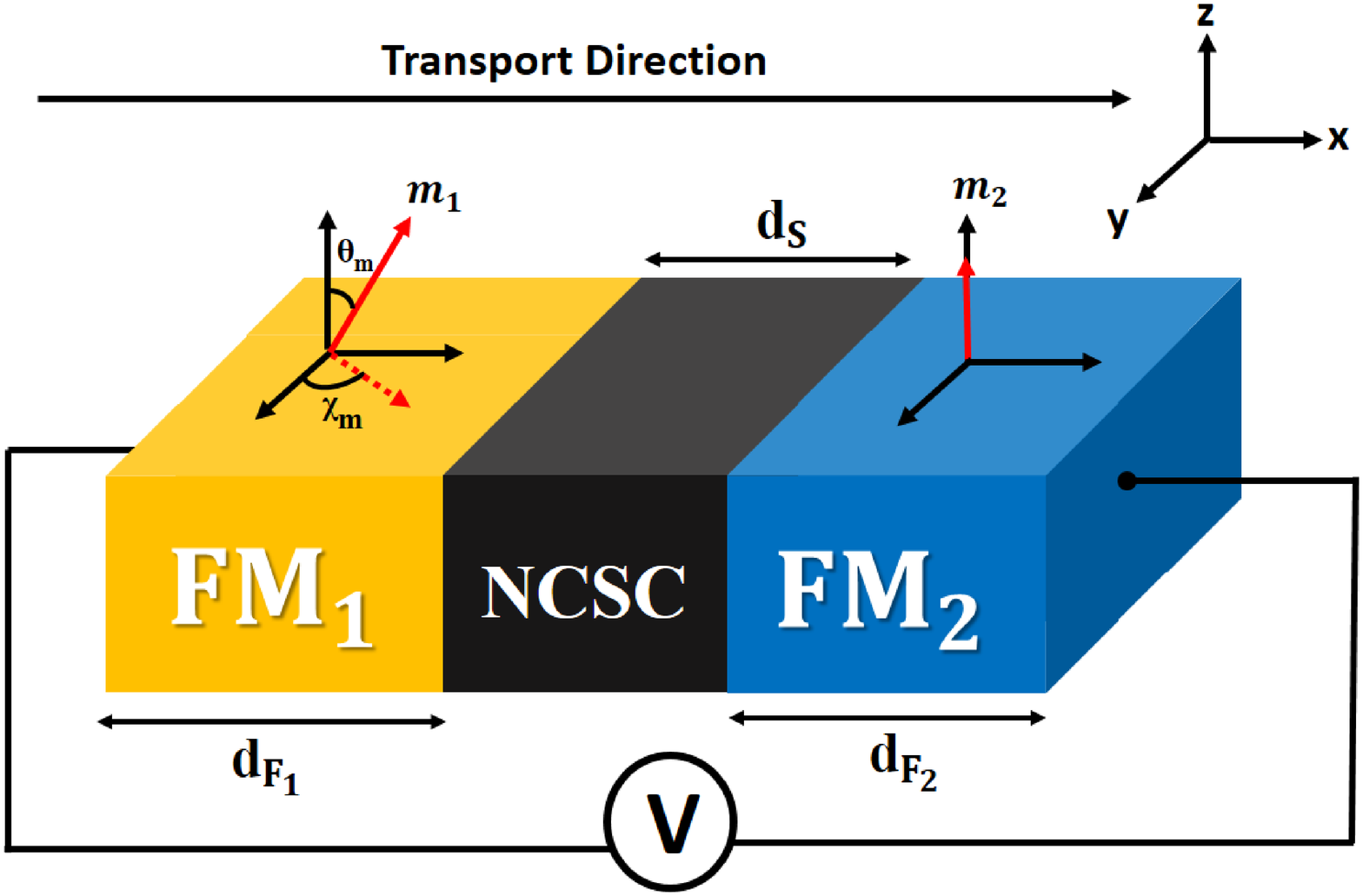}
\vspace{0.1cm}
\includegraphics[scale=0.27]{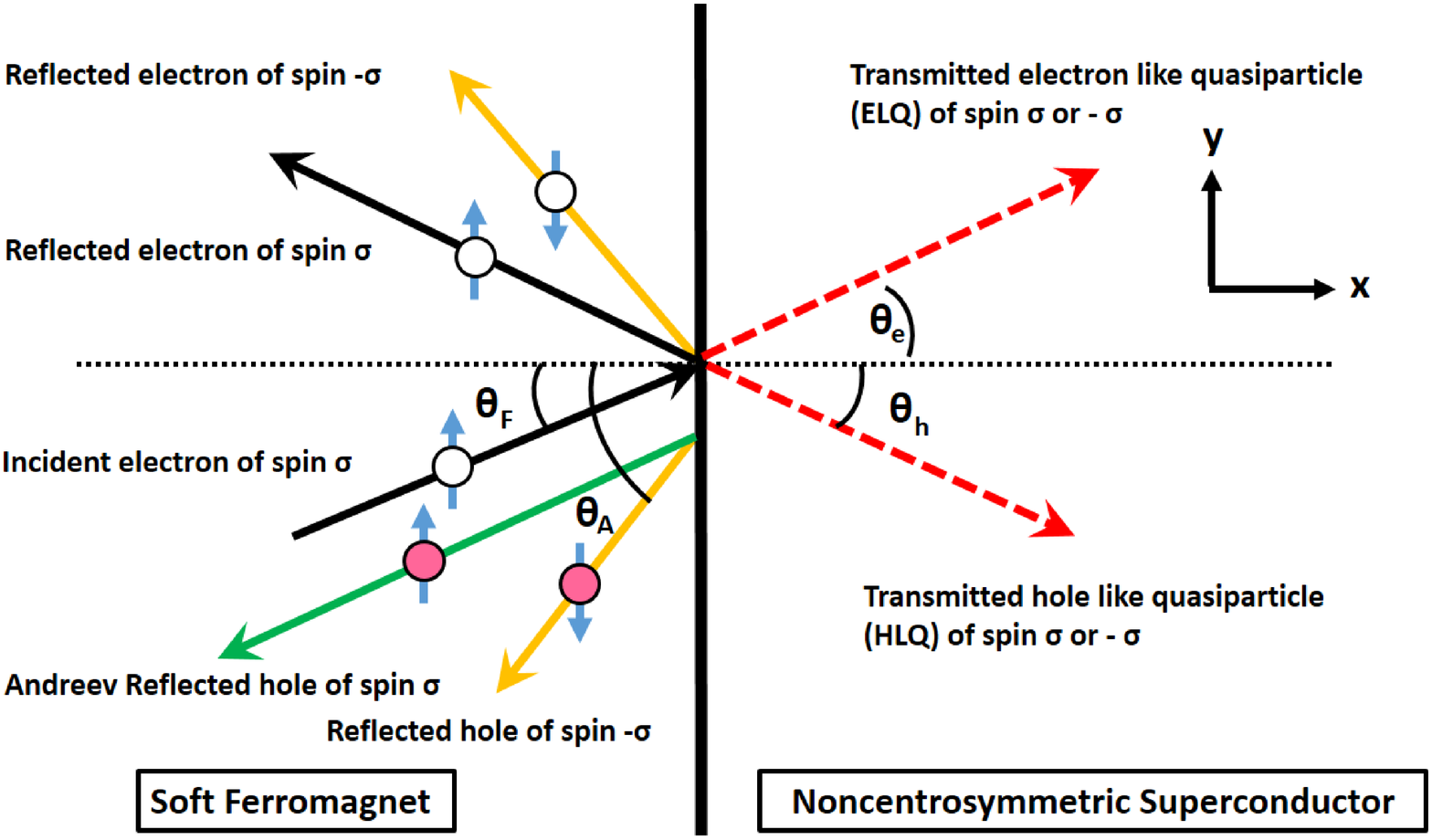}
}
\caption{The proposed experimental setup consists of noncentrosymmetric 
superconductor (NCSC) sandwitched between a soft and a hard ferromagnetic 
(FM) layers. The magnetization orientation of the soft and the hard 
FM layers are supposed to be misaligned by a polar angle $\theta_m$ and 
azimuthal angle $\chi_m$ as shown in the figure at top. The figure in the 
bottom gives a schematic overview of the scattering process takes place at the 
Ferromagnet$|$Noncentrosymmetric Superconductor (F$|$NCSC) junction.
When an electron with spin up in the free FM region is incident on the F$|$NCSC
interface there will be three possibilities: (i) it get reflected as a normal 
spin up electron, (ii) it gets retro-reflected as a hole with spin up or down 
configuration but in the opposite direction with the incident electron, and 
(iii) it gets transmitted in the S-region as a electron or hole like quasi 
particle with up or down spin. Here, we assumed x-direction as the direction 
of transport. Different colours of the ferromagnetic layers indicates the 
level of magnetization.}
\label{fig1}
\end{figure}

We consider a F$|$S$|$F spin valve shown in Fig.\ref{fig1} consisting of 
a NCSC layer sandwiched between two ferromagnetic layers having different 
strength 
of magnetization. The left ferromagnetic layer is soft and can have any 
arbitrary orientation of magnetization, while the magnetization  
of the right ferromagnetic layer is fixed. In our analysis we choose that the 
orientation of the hard ferromagnetic layer is along the z-direction.  We 
construct the Bogoliubov de-Gennes (BdG) Hamiltonian for the proposed setup and 
then solved the BdG equation at different interfaces to obtain the wave 
functions. The reflection and the transmission coefficients were then obtained 
by the proper choice of boundary conditions. In order to calculate the 
tunneling conductance at the F$|$NCSC interface of the F$|$S$|$F spin valve 
we use scattering matrix formalism to calculate the reflection and transmission 
coefficients and then by using an extended Blonder - Tinkham - Klapwijk (BTK) 
approach \cite{blonder}, we calculated the tunneling conductance
at the F$|$NCSC junction of the valve. 

The starting point of our theory is the Bogoliubov-de Gennes (BdG) equations, which can be read as
 \begin{equation}
 \label{eq1}
 \mathcal{H}_{BdG}\Psi_{\alpha}(r)= E\Psi_\alpha(r),
 \end{equation}
where $\Psi_{\alpha}(r)$ is the Bogoliubov wave function has the following 
form:
\begin{equation}
\label{eq2}
\Psi_{\alpha}(r) = (u_{n\uparrow} , u_{n\downarrow} , v_{n\uparrow}, v_{n\downarrow})^T. 
\end{equation}
Here, $u_{n}$s and $v_{n}$s are basis functions representing the spin-up 
($\uparrow$) and spin-down ($\downarrow$) states of electronlike and holelike 
quasiparticles respectively.

In matrix form the BdG Hamiltonian of the system can be
written as
\begin{equation}
\label{eq3}
\mathcal{H}_{BdG} = \left(
\begin{array}{cc}
 \hat{H}_0 & \hat{\Delta }_{\alpha \beta }(r,r') \\
 \hat{\Delta }_{\alpha \beta }^{\dagger }(r,r') & -\hat{H}_0^{\dagger } \\
\end{array}
\right)_{4\times 4},
\end{equation}
where the hat sign represents $2\times2$ matrices in spin space and $\hat{H}_0$
is the single particle Hamiltonian can be written as
\begin{equation}
\label{eq4}
\hat{H}_0 = \Big(-\frac{\nabla^2}{2} - E_{Fi} + U_{int}\Big)\hat{I} - \vec{h}\cdot\hat{\sigma}, 
\end{equation}
where $\hat{I}$ is the $2\times2$ unit matrix and $\hat{\sigma}$ are the
Pauli matrices. The first term appearing in this equation represents the single 
particle kinetic energy, where for simplicity we consider $\hbar = 1$ and 
the electronic mass m as unity. $E_{Fi}$ gives the Fermi energies in the 
respective layers. The ratio of square root of the Fermi energy in the 
superconducting layer $E_{FS}$ to the  Ferromagnetic layer $E_{FF}$ is denoted 
by a dimensionless parameter $\lambda$, which physically  represents Fermi 
Wave-vectors Mismatch (FWM) in the respective layers. That is, we define 
$\lambda = \sqrt{\frac{E_{FS}}{E_{FF}}} = \frac{q_{FS}}{k_{FF}}$, where
$q_{FS}$ and $k_{FF}$ are the Fermi momentum in superconducting and 
ferromagnetic layers respectively. $U_{int}$ gives the interaction potential 
of the F$|$NCSC interface located at $x = 0$ along the y direction.
To study the effect of RSOC, we consider our interaction potential $U_{int}$ 
as follows \cite{wu,wu1,cheng1,kapri}:
\begin{equation}
\label{eq5}
U_{int}(x) = \big[U_0 + U_R \hat{e}_x.(\hat{\sigma}\times\vec{k})\big]\delta(x),
\end{equation}
where $U_0$ gives the strength of spin independent potential, while $U_R$ 
characterize the strength of RSOC. $\hat{e}_x$ is an unit vector directed 
normal to the interface and $\vec{k} = -i\nabla$. 

The fourth term in Eq.(\ref{eq4}) represents exchange interaction. The 
exchange field $\vec{h}$ arises due to the ferromagnetic magnetization for the 
free layer is described by $\vec{h} =  h_0(\sin\theta_m\cos\chi_m,\; \sin\theta_m\sin\chi_m,\;\cos\theta_m)$, where $\theta_m$ and $\chi_m$ respectively are 
the polar and azimuthal angles of magnetization. 

The gap matrix $\hat{\Delta }_{\alpha \beta }(r,r')$ appearing in 
Eq.(\ref{eq3}) has the following form \cite{linder}:
\begin{equation}
\label{eq6}
 \hat{\Delta }_{\alpha \beta }(r,r') =
\left(
\begin{array}{cc}
 \Delta _{\uparrow\uparrow}(r,r') & \Delta _{\uparrow\downarrow}(r,r') \\
 \Delta _{\downarrow\uparrow}(r,r') & \Delta _{\downarrow\downarrow}(r,r') \\
\end{array}
\right)
\end{equation}
It should be noted that $\Delta _{\uparrow\downarrow}(r,r')$ is a superposition
of the singlet (S) and the triplet (T) components that satisfies 
\begin{align}
\Delta _{\uparrow\downarrow}(r,r')& = \Delta^S _{k\uparrow\downarrow}(r,r') +
 \Delta^T _{k\uparrow\downarrow}(r,r'),\\
\Delta^T _{k\uparrow\downarrow}(r,r')& = \Delta^T _{k\downarrow\uparrow}(r,r'),\\
\Delta^S _{k\uparrow\downarrow}(r,r')& = -\Delta^S _{k\downarrow\uparrow}(r,r').
\end{align}
Thus, in view of Eqs.(\ref{eq3}, \ref{eq4}, \ref{eq5}, \ref{eq6}) the BdG Hamiltonian can be written as
\begin{widetext}
 \begin{equation}
\label{eq7}
\mathcal{H}_{BdG} = \left(
\begin{array}{cccc}
 -h_z+H' & g_{k_-}-h_{\text{xy}} & \Delta^T _{k\uparrow\uparrow} & \Delta^S _{k\uparrow\downarrow}+\Delta^T _{k\uparrow\downarrow} \\
 g_{k_+}-h^*{}_{\text{xy}} & h_z+H' & -\Delta^S _{k\uparrow\downarrow}+\Delta^T _{k\uparrow\downarrow} & \Delta^T _{k\downarrow\downarrow} \\
 {\Delta^T_{k\uparrow\uparrow}}^\dagger &  -{\Delta^S _{k\uparrow\downarrow}}^\dagger + {\Delta^T _{k\uparrow\downarrow}}^\dagger & h_z-H' & g_{k_+}-h^*{}_{\text{xy}} \\
 {\Delta^S _{k\uparrow\downarrow}}^\dagger+{\Delta^T _{k\uparrow\downarrow}}^\dagger &  {\Delta^T_{k\downarrow\downarrow}}^\dagger & g_{k_-}-h_{\text{xy}} & -h_z-H' \\
\end{array}
\right),
  \end{equation}
\end{widetext}
where $h_{xy} = h_x - i h_y$, $g_{k\pm} = U_R(k_x \pm i k_y)\Theta(x)$ and 
$H' = -\frac{\nabla^2}{2} - E_{Fi} + U_0\delta(x)$. $\Theta(x)$ is the 
Heavyside step function defined by,
\begin{equation}
\label{eq7a}
\Theta(x) = \left\{\begin{array}{rc} 
0, & x <0,\\
1, & x \ge 0.
\end{array} \right.
\end{equation}

Diagonalizing the Hamiltonian in Eq.(\ref{eq7}), we obtain the momenta of 
electrons and holes in the different regions, which is necessary to calculate 
the normal and Andreev reflection coefficients, which will be further 
necessary to calculate the tunneling conductance in a F$|$NCSC junction. 
Denoting the momenta for the electrons and the holes in the ferromagnetic 
region respectively as $k^+$ and $k^-$, we get after diagonalizing the
BdG Hamiltonian appearing in Eq.(\ref{eq7}) as
\begin{equation}
\label{eq8}
k^{\sigma} = \sqrt{2(E_{FF} + U_{int} + \vec{h}.\vec{\sigma} \pm E)},
\end{equation}
where $\sigma = \pm 1$. In a
 more explicit way the Eq.(\ref{eq8}) can be written as
\begin{equation}
\label{eq9}
k^{\sigma} =  k_{FF}\sqrt{1 + Z_0 -\sigma Z_R \sin\theta_F -\sigma X \pm Z_1}, 
\end{equation}
where for simplicity we define $Z_0 = \frac{2U_0}{k_{FF}}$, $Z_R = 2U_R$, 
$X = \frac{M}{E_{FF}}$ and $Z_1 = \frac{E}{E_{FF}}$. Here $M$ is the 
magnetization of the feromagnetic region and hence $X$ gives the magnetization
strength per unit Fermi energy of the region. Furthermore, the momenta of the 
electron-like and the hole-like quasiparticles in the superconducting region 
respectively are denoted by $q^+$ and $q^-$, which are found as
\begin{equation}
\label{eq10}
q^\pm = \sqrt{2(E_{FS}\pm\sqrt{E^2-\Delta_{\alpha\beta}^2})}.
\end{equation}

In order to make the calculation easier we set an approximation: 
$q^+ = q^- = q_{FS}$, where $q_{FS}$ is the Fermi momentum in the 
superconducting region. The approximation yields an error of the order 
of $\frac{\delta q_{FS}}{q_{FS}} = \frac{\sqrt{E^2 - \Delta_{\alpha\beta}^2}}{E_{FS}}$, which is of the order of $\frac{\Delta_{\alpha\beta}}{E_{FS}}$ since, 
$\Delta_{\alpha\beta} \ll E_{FS}$, hence it is a valid  approximation.

Again, in the tunneling process the parallel component of momenta is conserved. So we can write,
\begin{equation}
\label{eq11}
k^+\sin\theta_F = k^-\sin\theta_A = q^+\sin_{\theta_e} = q^-\sin\theta_h,
\end{equation}
where $\theta_F$ and $\theta_A$ are the angle of incidence of the electron in 
ferromagnetic region and the retro reflected angle of the hole in the 
superconducting region respectively. $\theta_e$ is the angle of refraction
for the electron like quasiparticles, while $\theta_h$ is the angle of 
refraction for the hole like quasiparticles.
 
Choosing a plane wave solution $\Psi(x) = \psi(x) e^{i k x}$, and introducing 
it in the Eq.(\ref{eq1}) we obtain the wave function for the ferromagnetic 
region $\Psi_{\text{FM}}(x)$ with arbitrary orientation of magnetization. It 
can be written as
\begin{multline}
\label{eq12}
\Psi_{\text{FM}}(x) = s_{\uparrow }\left(
\begin{array}{c}
\cos\theta_m  \\
\sin\theta_m  e^{-i \chi_m } \\
0 \\
0 \\
\end{array}
\right)e^{i k^{+}\cos\theta_Fx}
\\+s_{\downarrow }\left(
\begin{array}{c}
-\sin\theta_m e^{i \chi_m } \\
\cos\theta_m \\
0 \\
0 \\
\end{array}
\right)e^{ik^{-}\cos\theta_Fx}
\\+r_e^{\uparrow }\left(
\begin{array}{c}
\cos\theta_m  \\
\sin\theta_m e^{-i \chi_m } \\
0 \\
0 \\
\end{array}
\right)e^{-i k^{+}S_1x}
\\+r_e^{\downarrow }\left(
\begin{array}{c}
-\sin\theta_m e^{i\chi_m } \\
\cos\theta_m \\
0 \\
0 \\
\end{array}
\right)e^{-i k^{+}S_2x}
\\+r_h^{\uparrow }\left(
\begin{array}{c}
0 \\
0 \\
\cos\theta_m \\
\sin\theta_m e^{-i \chi_m } \\
\end{array}
\right)e^{ik^{+}S_1x}
\\+r_h^{\downarrow }\left(
\begin{array}{c}
0 \\
0 \\
-\sin\theta_m e^{i \chi_m } \\
\cos\theta_m\\
\end{array}
\right)e^{ik^{-}S_2x},
\end{multline}
where $S_1 = s_\uparrow cos\theta_{\textbf{F}} + s_\downarrow cos\theta_{\textbf{A}}$ 
and $S_2 = s_\uparrow cos\theta_{\textbf{A}} +  s_\downarrow cos\theta_{\textbf{F}}$. For up spin incident particle 
we choose $s_\uparrow = 1$, $s_\downarrow = 0$, while for a down spin 
particle $s_\uparrow = 0$ , $s_\downarrow = 1$. $\theta_m$ and $\chi_m$ 
respectively represents the polar angle of magnetization and the azimuthal 
angle of magnetization as shown in Fig.(\ref{fig1}). $r_e^{\uparrow }$ 
($r_e^{\downarrow }$ ) is the normal reflection coefficient for upspin 
(downspin) electron, while $r_h^{\uparrow }$ ($r_h^{\downarrow }$) is the 
retro reflection coefficient for the upspin (downspin) hole.  

In a similar way for the superconducting layer the wave function can be 
written \cite{linder} as
\begin{multline}
\label{eq13}
\Psi _{\text{SC}}(x)=\frac{t_e{}^{\uparrow }}{\sqrt{2}}\left(
\begin{array}{c}
 u_+ \\
 u_+ e^{-i \phi } \\
 -v_+ e^{-i \phi } \\
 v_+ \\
\end{array}
\right)e^{iq_e^{+}\cos \theta _ex}
\\+\frac{t_e{}^{\downarrow }}{\sqrt{2}}\left(
\begin{array}{c}
 u_- \\
 -u_- e^{-i \phi } \\
 v_- e^{-i \phi } \\
 v_- \\
\end{array}
\right)e^{iq_e^{-}\cos \theta _ex}
\\+\frac{t_h{}^{\uparrow }}{\sqrt{2}}\left(
\begin{array}{c}
 v_+ \\
 v_+ e^{-i \phi } \\
 -u_+ e^{-i \phi } \\
 u_+ \\
\end{array}
\right)e^{iq_h^{+}\cos \theta _hx} 
\\+\frac{t_h{}^{\downarrow }}{\sqrt{2}}\left(
\begin{array}{c}
 v_- \\
 -v_- e^{-i \phi } \\
 u_- e^{-i \phi } \\
 u_- \\
\end{array}
\right)e^{i\text q_h^{-}\cos \theta _hx},
\end{multline}
where $\phi$ is the superconducting phase factor, $t_e^{\uparrow }$( $t_e^{\downarrow }$) corresponds to the transmission coefficient for up(down) spin 
Electron-Like Quasiparticles (ELQs), while $t_h^{\uparrow }$( $t_h^{\downarrow}$) represents the transmission coefficients for up(down) spin Hole-Like  
Quasiparticles (HLQs). The amplitudes of  ELQs and HLQs are given by,
\begin{eqnarray}
\label{eq14}
\label{eq15}
 u_\pm =  \frac{1}{\sqrt{2}}\sqrt{1+\frac{\Omega_{\pm}}{E}},\\
 v_\pm =  \frac{1}{\sqrt{2}}\sqrt{1-\frac{\Omega_{\pm}}{E}},
\end{eqnarray}
with $\Omega_{\pm}$ are considered to be as
\begin{equation}
\label{eq16}
\Omega_{\pm} = \sqrt{E^2-|\Delta_s\pm\frac{\Delta_t}{2}|^2}.
\end{equation}
Here, $\Delta_s$ and $\Delta_t$ are singlet and triplet gap parameters 
respectively of superconducting states.

The wave functions $\Psi_\text{FM}(x)$ and $ \Psi_\text{SC}(x)$ must satisfy 
the following boundary conditions:
\begin{align}
\label{eq17}
\Psi_\text{FM}(x = 0^-)& = \Psi_\text{SC}(x = 0^+),\\
\label{eq18}
\partial_x[\Psi_\text{SC}(x = 0^+)-\Psi_\text{FM}(x = 0^-)]& = 2U_{int}\Psi_\text{FM}(x=0).
\end{align}
The reflection coefficients ($r_e^\sigma $, $r_h^\sigma$) and the transmission 
coefficients ($t_e^\sigma $, $t_h^\sigma$) are calculated using these boundary 
conditions. Though the explicit expressions for the reflection and the 
transmission probabilities are too large, however, we have presented
an analytic form of the normal and Andreev reflection coefficients for the 
incidence of an upspin electron in the Appendix.

\subsection{Conductance spectra at the F$|$NCSC junction}
The tunneling conductance of our setup can be calculated by using 
Blonder - Tinkham - Klapwijk (BTK) formalism \cite{blonder}. The normalized
tunneling conductance $G^\sigma(E,\theta_F)$ for an incoming electron of spin 
$\sigma$ with an incident angle $\theta_F$ at the junction reads as
\begin{equation}
\label{eq19}
G^\sigma(E,\theta_F) = G_N^{-1}\big[1 + \sum_\sigma(|r_h^\sigma (E,\theta_F)|^2 - |r_e^\sigma(E,\theta_F)|^2)\big],
\end{equation}
where $G_N$ is the tunneling conductance for N$|$N (Normal metal-Normal metal)
junction for an interface potential $U_0$ and has the following form:
\begin{equation}
\label{eq20}
G_N = \int_{-\frac{\pi}{2}}^{\frac{\pi}{2}} d\theta_F \frac{4\cos^3\theta_F}{4\cos^2\theta_F + Z_0^2}.
\end{equation}
Thus, in view of this the angularly averaged conductance can be written as \cite{kapri,cheng1,blonder,linder,linder1}
\begin{equation}
\label{eq21}
G(E) = G_N^{-1} \int_{-\frac{\pi}{2}}^{\frac{\pi}{2}}  d\theta_F \cos\theta_F P^\sigma G^\sigma(E,\theta_F),
\end{equation}
where $P^\sigma$ is the spin dependent probability factor for the spin 
injection $\sigma$ and can be read as \cite{linder1} 
\begin{equation}
\label{eq22}
P^\sigma = \frac{(1 + \sigma X)}{2}.
\end{equation}
\begin{figure*}[hbt]
\centerline
\centerline{
\includegraphics[scale = 0.57]{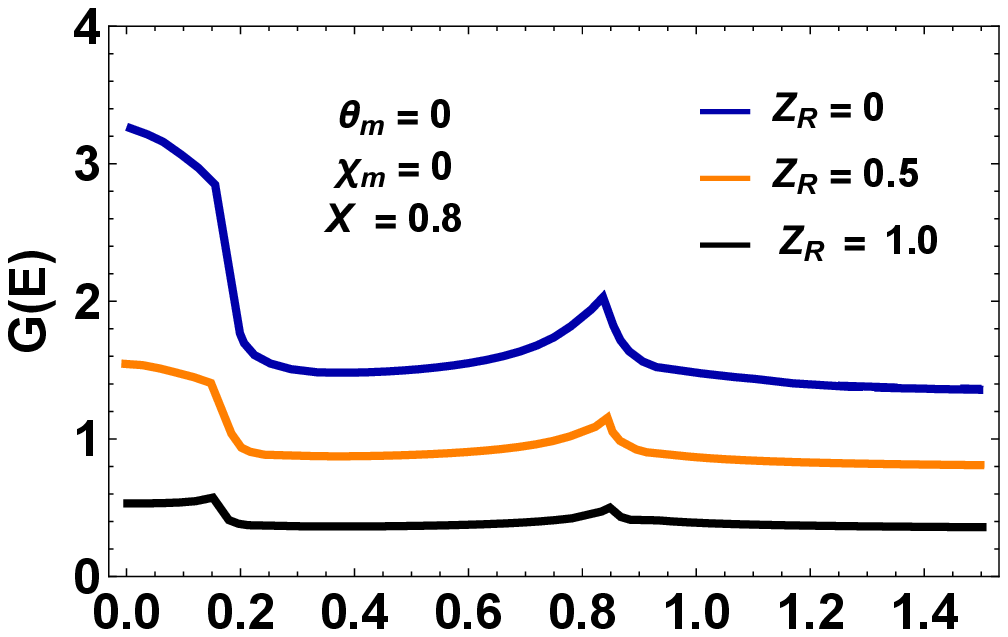}
\hspace{0.15cm}
\vspace{0.25cm}
\includegraphics[scale = 0.57]{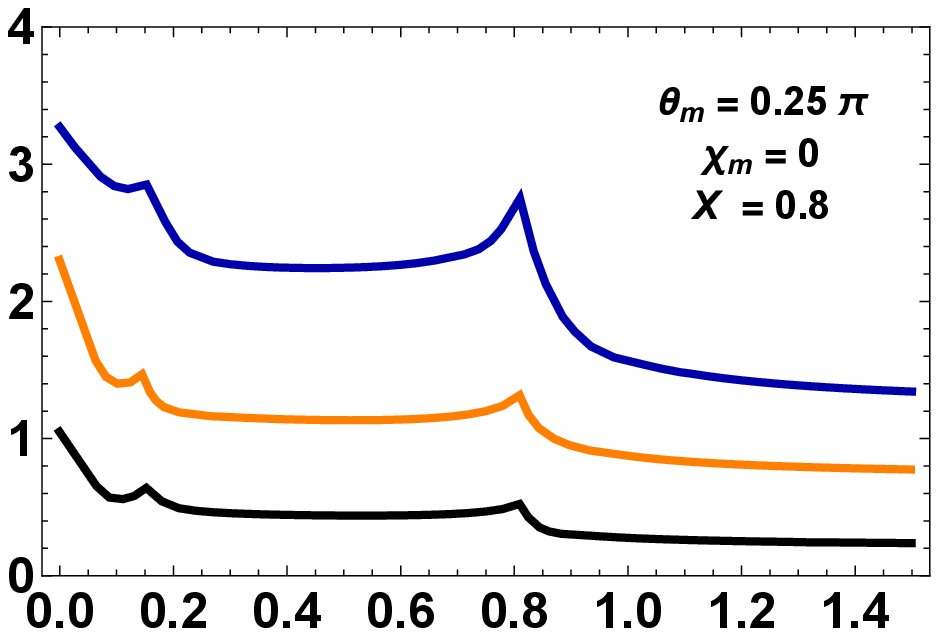}
\hspace{0.15cm}
\vspace{0.25cm}
\includegraphics[scale = 0.57]{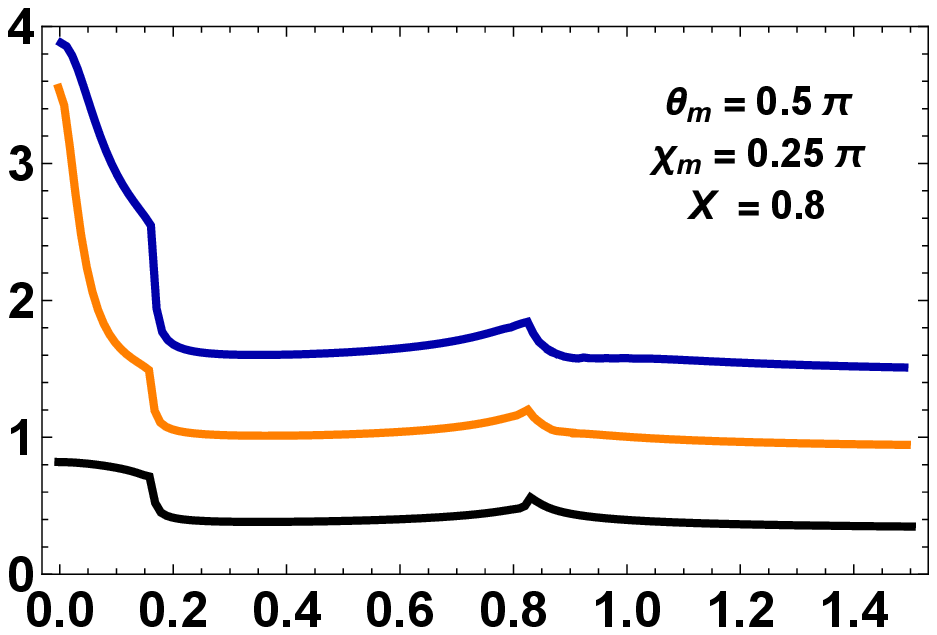}
\hspace{0.15cm}
\vspace{0.25cm}
\includegraphics[scale = 0.57]{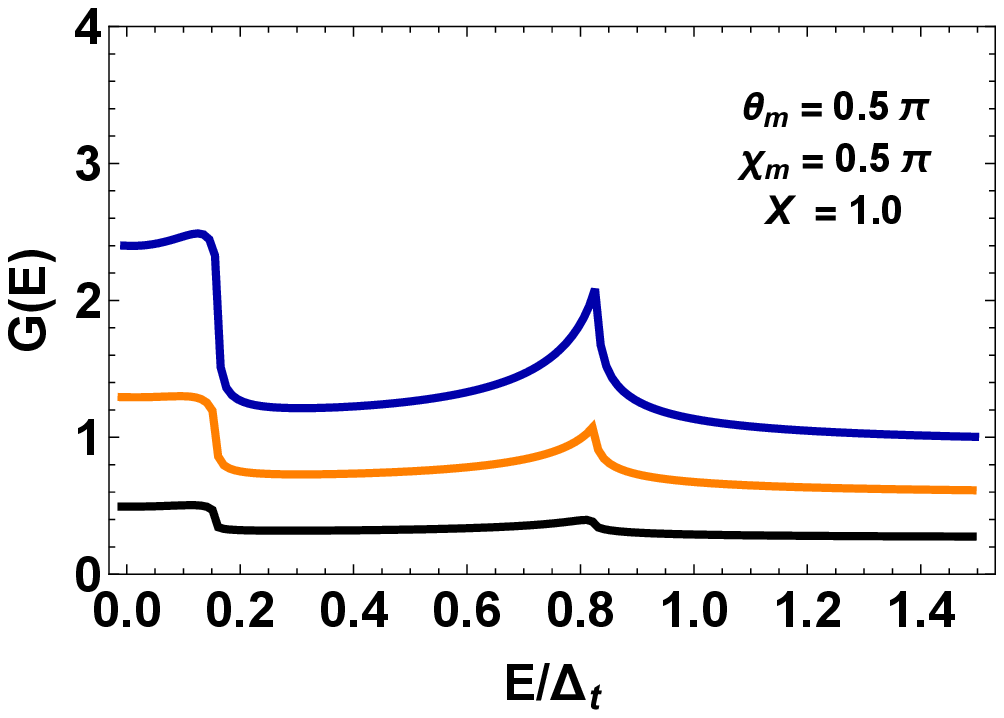}
\hspace{0.15cm}
\includegraphics[scale = 0.57]{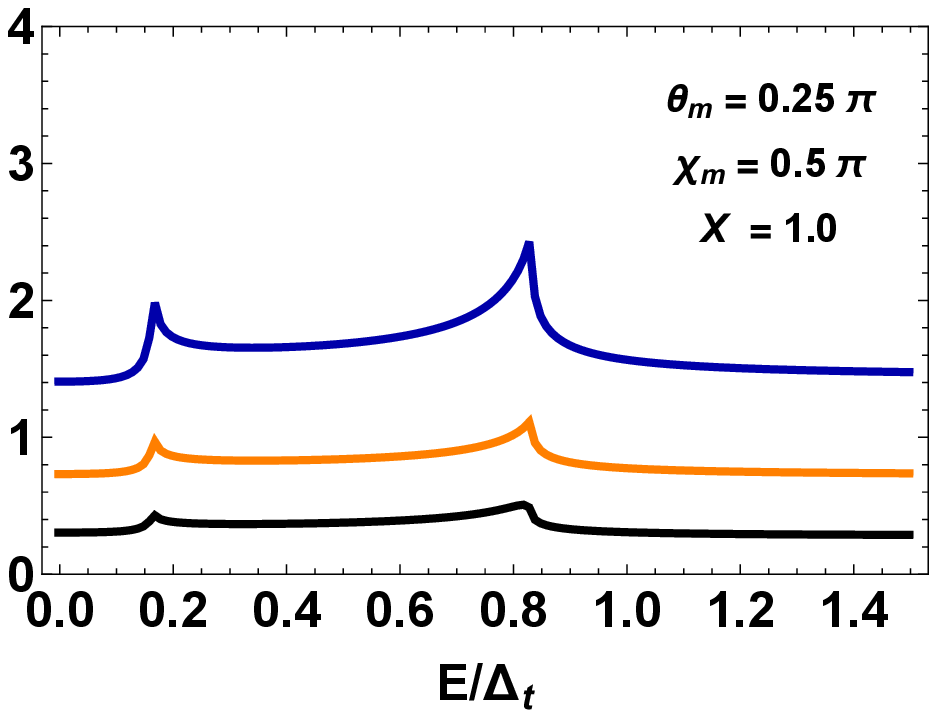}
\hspace{0.15cm}
\includegraphics[scale = 0.57]{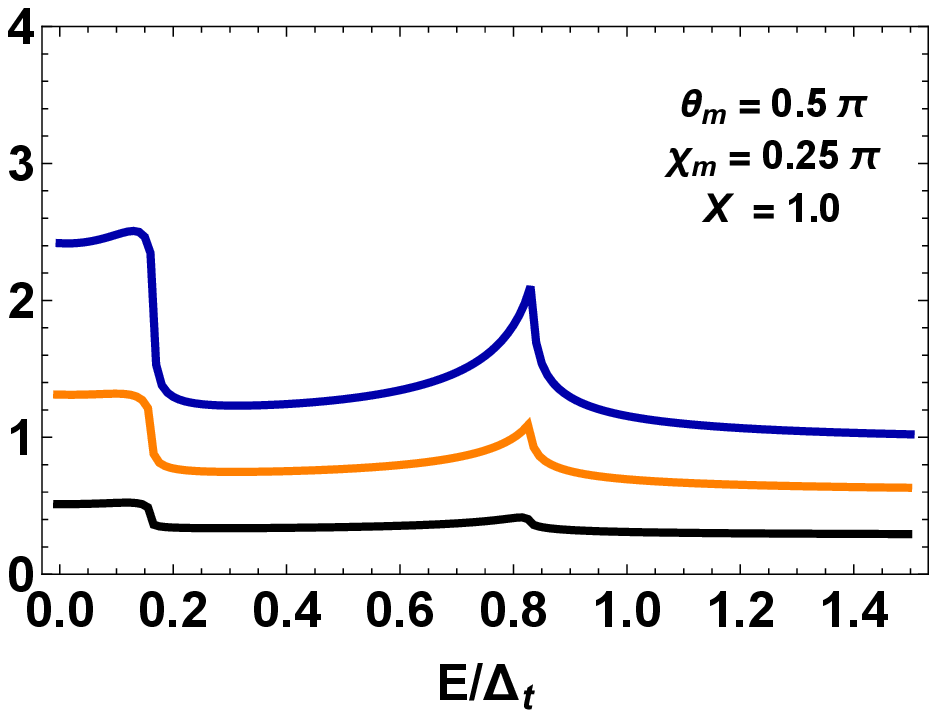}
\hspace{0.15cm}
}
\caption{Conductance spectra for different values of $Z_R$ with
$\Delta_s =\frac{\Delta_t}{3}$, $Z_0 = 0.1$, $\lambda = 1.5$. 
The figures in the top panel is for $X = 0.8$, while the bottom panel 
is for $X = 1.0$ for different orientations of magnetization.}
\label{fig2}
\end{figure*}
\begin{figure*}[hbt]
\centerline
\centerline{
\includegraphics[scale = 0.55]{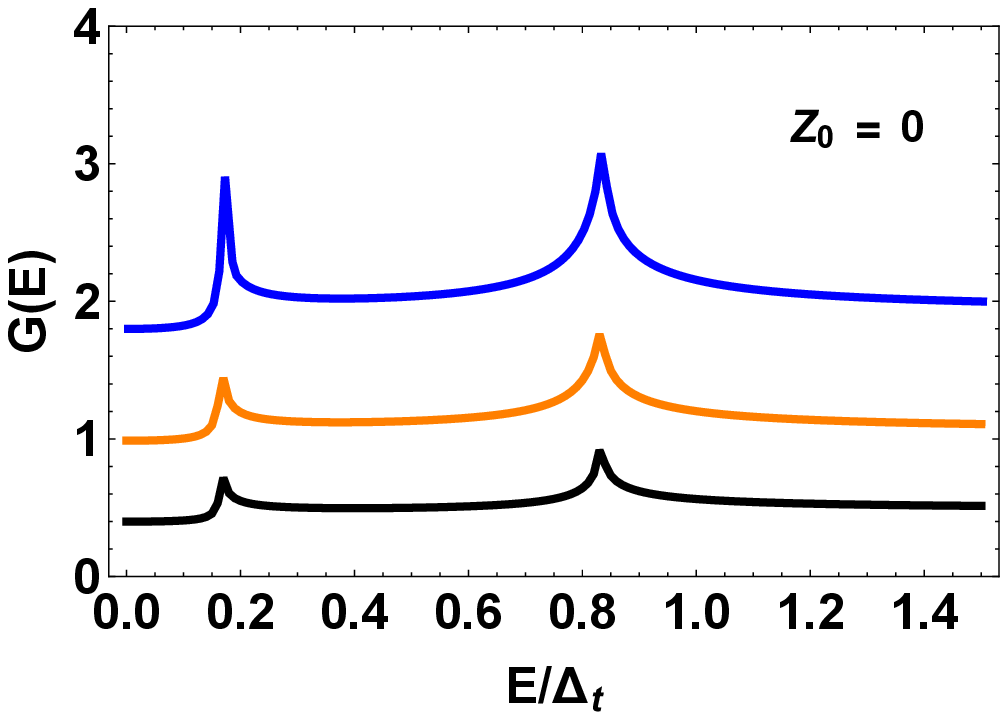}
\hspace{0.2cm}
\includegraphics[scale = 0.58]{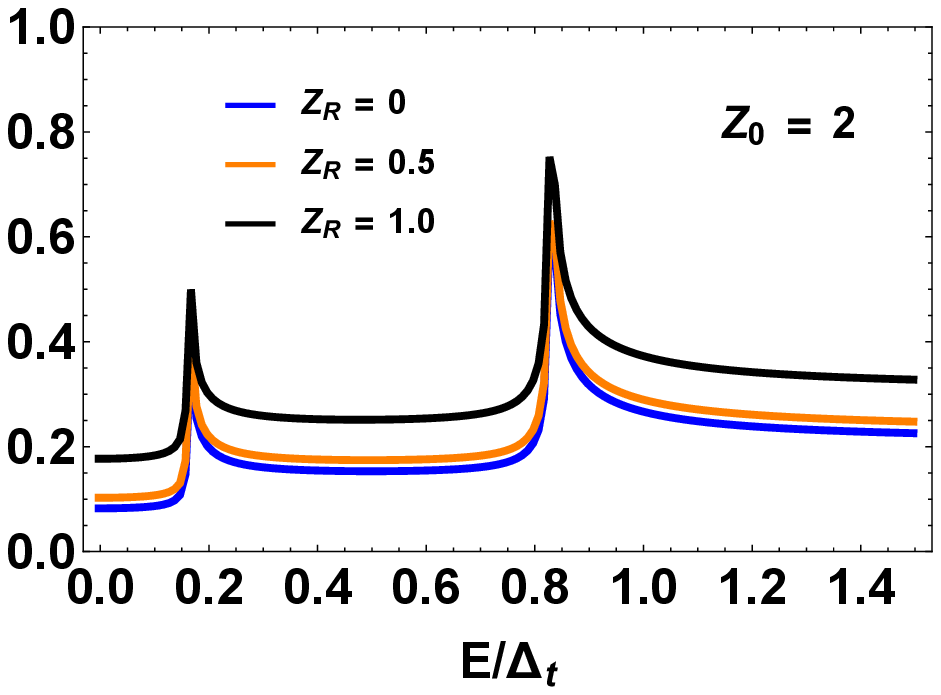}
\hspace{0.2cm}
\includegraphics[scale = 0.58]{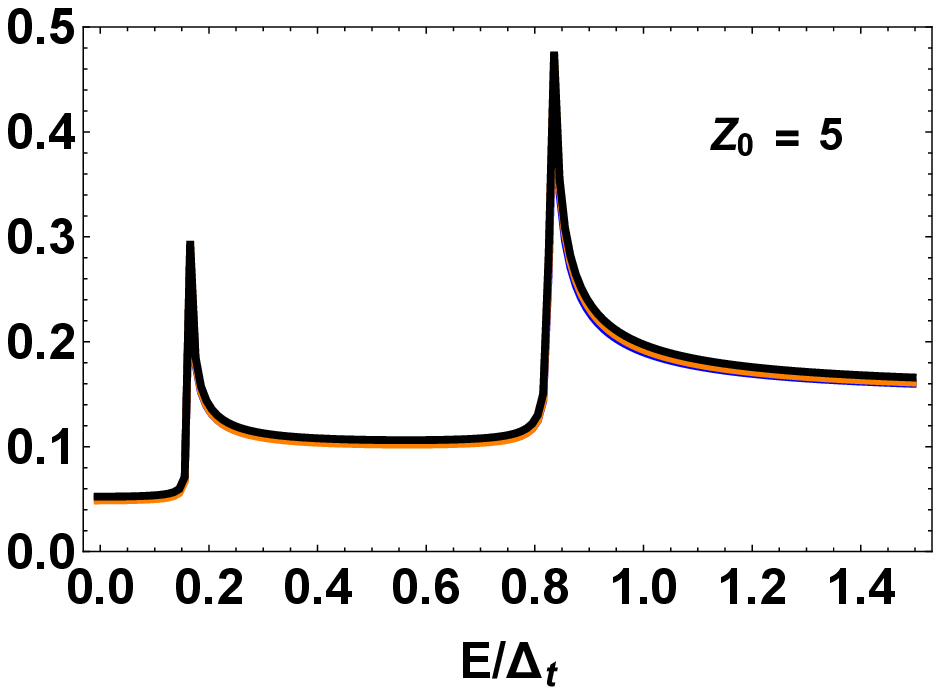}
\hspace{0.2cm}
}
\caption{Dependence of conductance $G(E)$ on the barrier transparency $Z_0$ for 
$\theta_m = 0.25\pi$, $\chi_m = 0.5\pi$ and $X = 0.8$ with $\lambda$ = $1.5$.}
\label{fig3}
\end{figure*}

\vspace{-1.0cm}
In this work, we have plotted the angularly averaged normalized conductance  
$G(E)$ that appears in (\ref{eq21}) as a function of biasing energy $E$ scaled 
by the gap amplitude parameter $|\Delta_\pm|$, where $|\Delta_\pm| = |\Delta_s\pm\frac{\Delta_t}{2}|$. From the earlier works \cite{linder1,bozovic1,bozovic2}, it is seen that the conductance has a strong dependence on 
singlet-triplet gap magnitudes. In all our analysis, we considered 
$\Delta_s =\frac{\Delta_t}{3}$. However, to understand its effect on 
conductance characteristics, we have also investigated 
the conductance spectrum for different singlet-triplet mixing magnitudes too,
since in NCSC the gap magnitude played a very important role. 
It is well known from the previously mentioned experimental works that the 
NCSC possess a strong ASOC \cite{gorkov,yip,samokhin,sergienko,frigeri,fujimoto1,fujimoto2} due to the lack of inversion center. Recently, with the increased 
applications of spintronics in last few years, the phenomenon of SOC gains a 
lot of attention. So, the central point of our work is to study the effect of 
RSOC on the tunneling conductance spectrum. 

\begin{figure*}[hbt]
\centerline
\centerline{
\includegraphics[scale = 0.65]{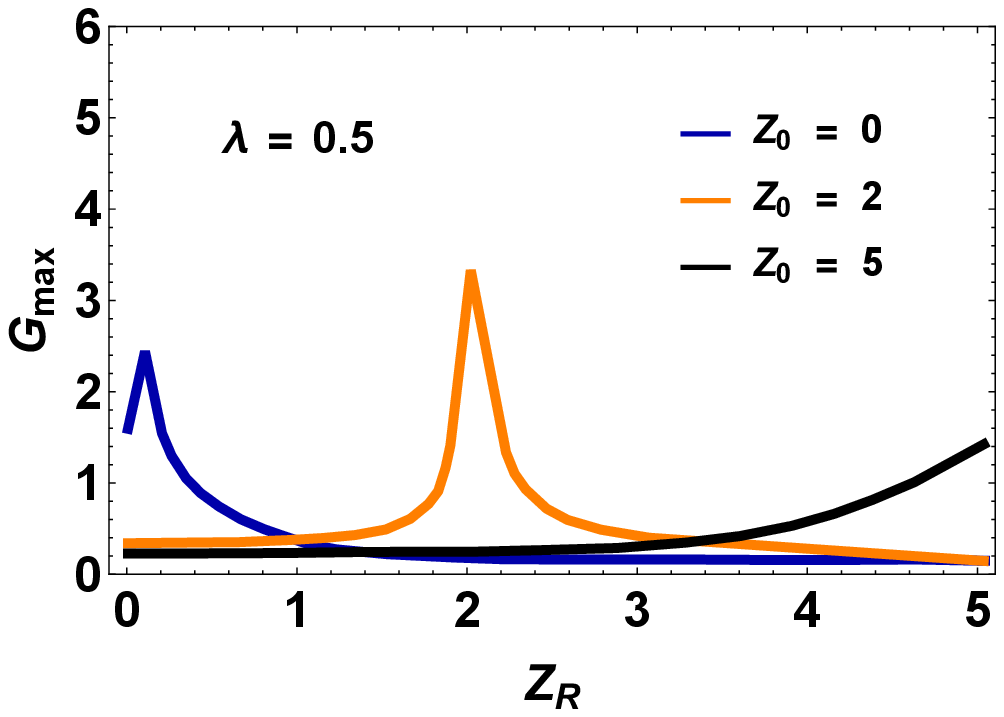}
\hspace{0.50cm}
\includegraphics[scale = 0.65]{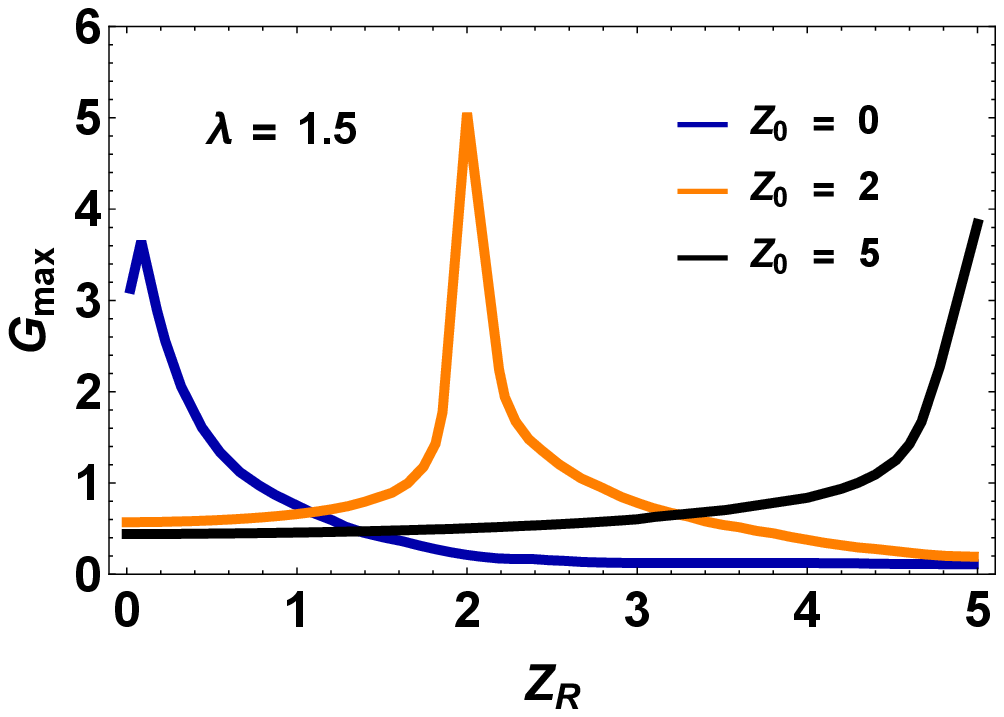}
\hspace{0.50cm}
}
\caption{Variation of G$_{max}$ with RSOC parameter $Z_R$ for different 
barrier strengths $Z_0 = 0 , 2$ and $5$ respectively. 
The plot in the left is for $\lambda = 0.5$, while the plot in the right is 
for $\lambda = 1.5$.}
\label{fig4}
\end{figure*}

Though magnetization plays a very important role in ferromagnetic 
superconductors, but its interplay with RSOC and hence on the conductance 
spectrum characteristics in F$|$NCSC junction of a F$|$S$|$F spin valve is 
still need to be understood. So we also investigated the role of magnetization 
strength ($X$), polar ($\theta_m$) and the azimuthal angles of magnetization 
($\chi_m$) on the tunneling conductance. Moreover, we have also investigated 
its effect on the RSOC. To understand the orientation dependence 
of the conductance characteristics, we have plotted the Zero Bias Conductance 
(ZBC) with polar angle of magnetization for different azimuthal angles of 
magnetization, different magnetization strength and also for different RSOC. 
Moreover, the dependence of ZBC on the azimuthal angles of magnetization is 
also studied for different magnetization strengths, polar angles of 
magnetization as well as for different RSOC's.

It should be noted here that there exist a strong dependence of the 
conductance characteristics on the barrier transparency. So we also investigated
the effect of barrier transparency on  conductance spectrum, RSOC as well as on 
magnetization. Usually, in Andreev reflection experiments of an F$|$NCSC 
junction, the Fermi momentum in the different regions are different. It arises 
due to unequal densities of the local charge carries in different regions. In 
this regard, the effect of FWM on conductance 
characteristics of an F$|$NCSC junction and how it will competes with 
magnetization and RSOC need to be understood properly, and hence we have 
considered a dimensionless parameter $\lambda$, which measures the ratio of 
Fermi wave-vectors in different layers as mentioned earlier. In this work, we 
consider the value of $\lambda$ as $0.5$, $1.0$, $1.25$ and $1.5$. For all our 
analysis we consider $Z_1$ to be very small and we assume $Z_1 = 0.001.$

\begin{figure*}[hbt]
\centerline
\centerline{
\includegraphics[scale = 0.6]{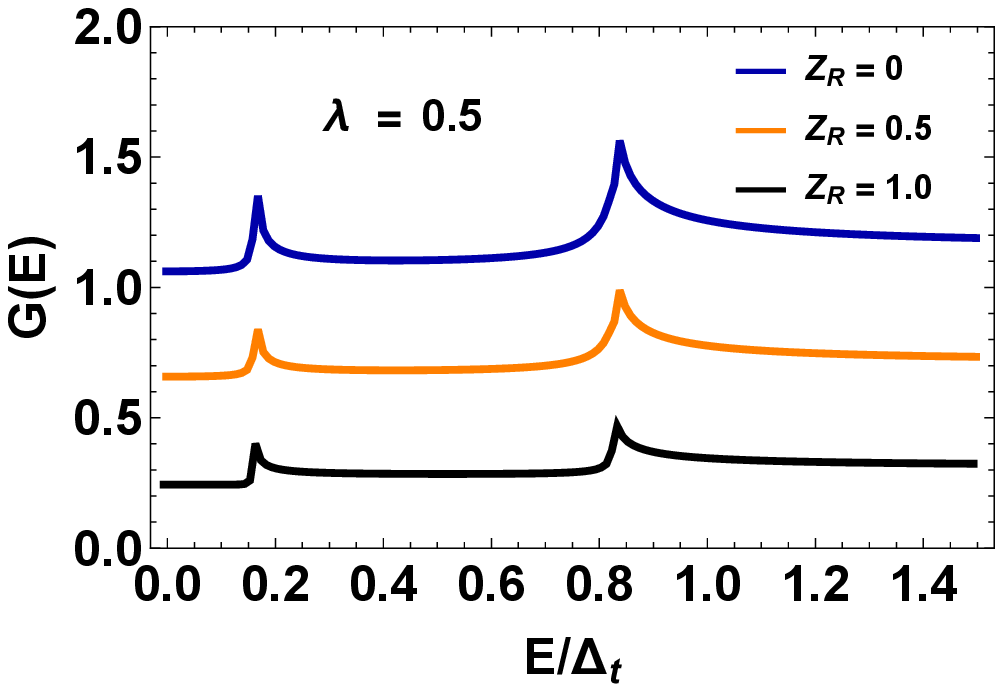}
\hspace{0.05cm}
\includegraphics[scale = 0.6]{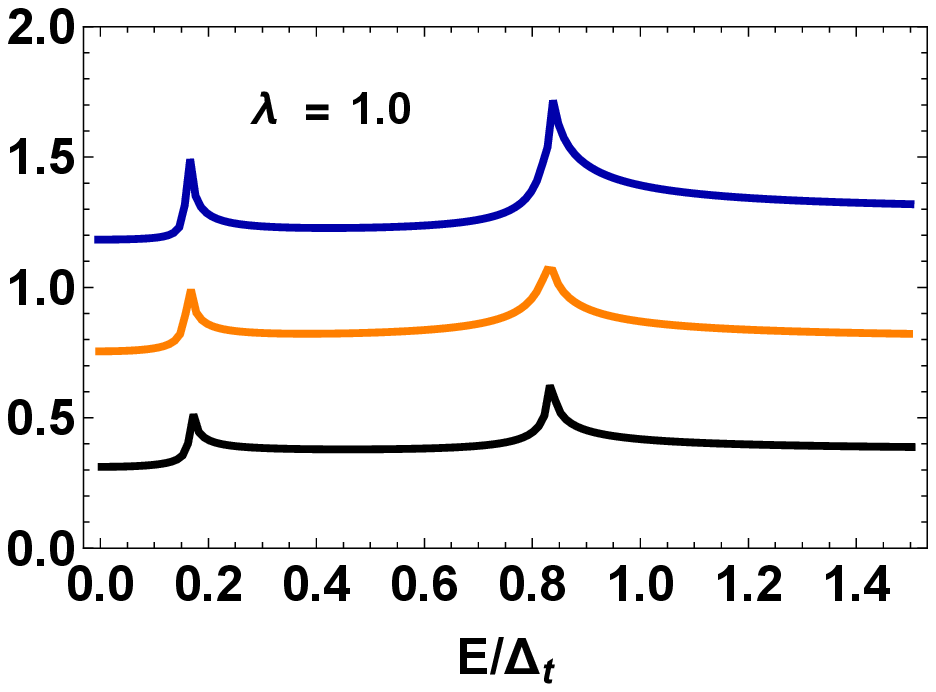}
\hspace{0.05cm}
\includegraphics[scale = 0.6]{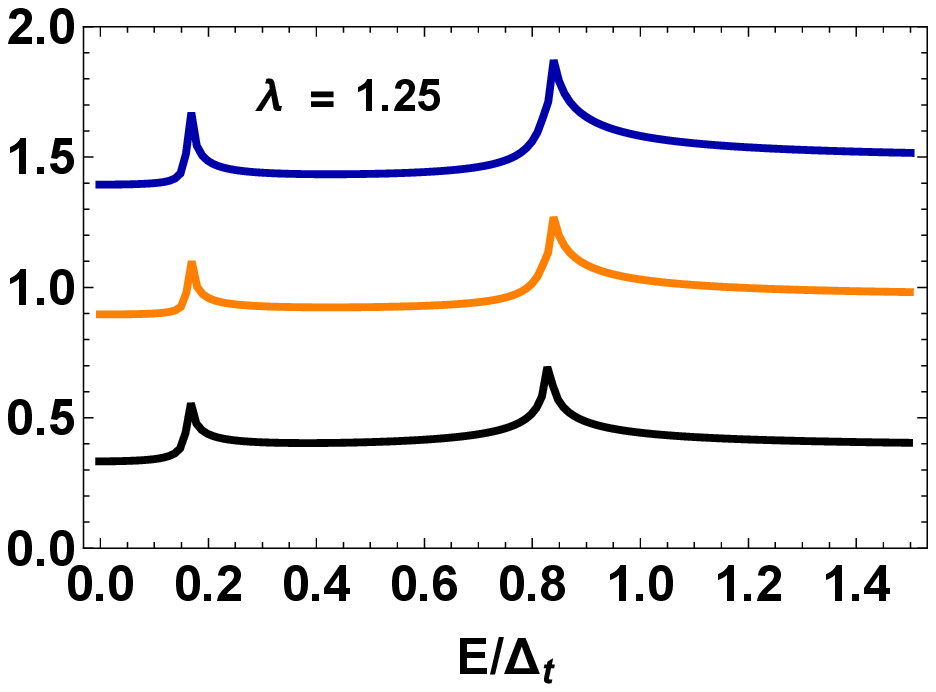}
\hspace{0.05cm}
}
\caption{Conductance spectra for different FWM parameter $\lambda$ with 
magnetization strength $X = 0.8$, polar angle of magnetization 
$\theta_m = 0.25\pi$, azimuthal angle of magnetization $\chi_m = 0.5\pi$ and 
barrier strength $Z_0 = 0$. The three different lines in the plots represent 
different choices of RSOC ($Z_R$).}
\label{fig5}
\end{figure*} 

\section{Results and Discussions}
\subsection{Effect of Rashba spin orbit coupling (RSOC)}
To understand the effect of RSOC on the tunneling conductance, we have 
plotted the normalized conductance $G(E)$ with biasing energy in 
Fig.(\ref{fig2}) for three different choices of RSOC strength, viz., 
$Z_R = 0, 0.5$ and $1.5$ considering a nearly transparent barrier 
($Z_0 = 0.1$). 
We consider the singlet-triplet gap amplitudes as $\Delta_s$ = 
$\frac{\Delta_t}{3}$ for our analysis. The figures in the top panel are 
plotted for a magnetization strength $X = 0.8$ 
while for the figures in the bottom panel we consider $X = 1.0$. The 
mismatch parameter $\lambda$ is considered to be as $1.5$ for our analysis in 
Fig.(\ref{fig2}). For the figures in the top panel we have considered the 
polar angle $\theta_m$ as $0, 0.25\pi$ and $0.5\pi$ keeping 
the azimuthal angle $\chi_m = 0$ in the first two plots of top panel of 
Fig.(\ref{fig2}). The figure on the right of top panel is drawn for $\theta_m = 0.5\pi$ and $\chi_m = 0.25\pi$.

\begin{figure*}[hbt]
\centerline
\centerline{
\includegraphics[scale = 0.6]{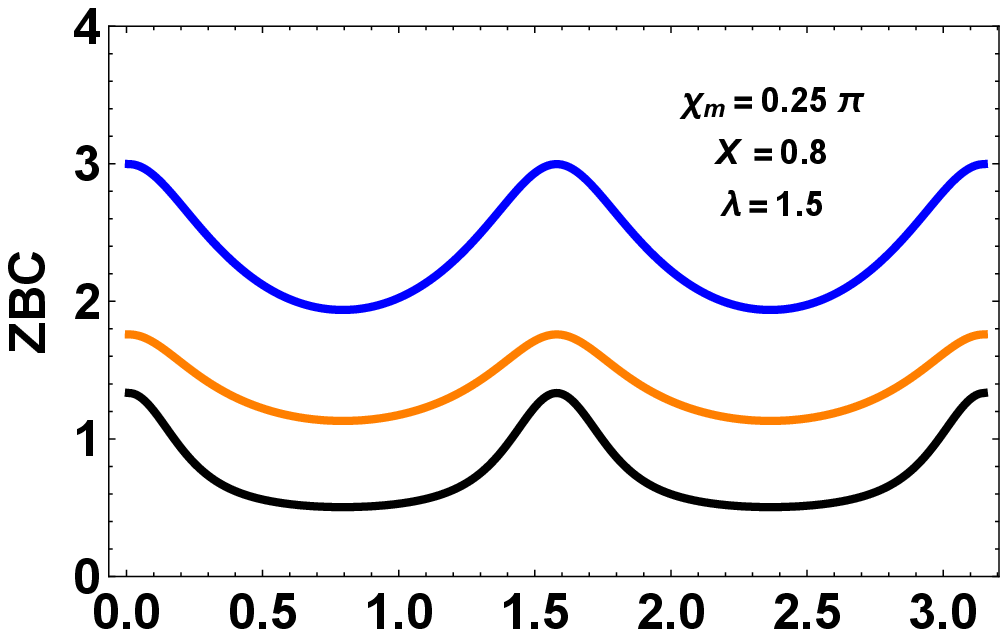}
\hspace{0.1cm}
\vspace{0.1cm}
\includegraphics[scale = 0.62]{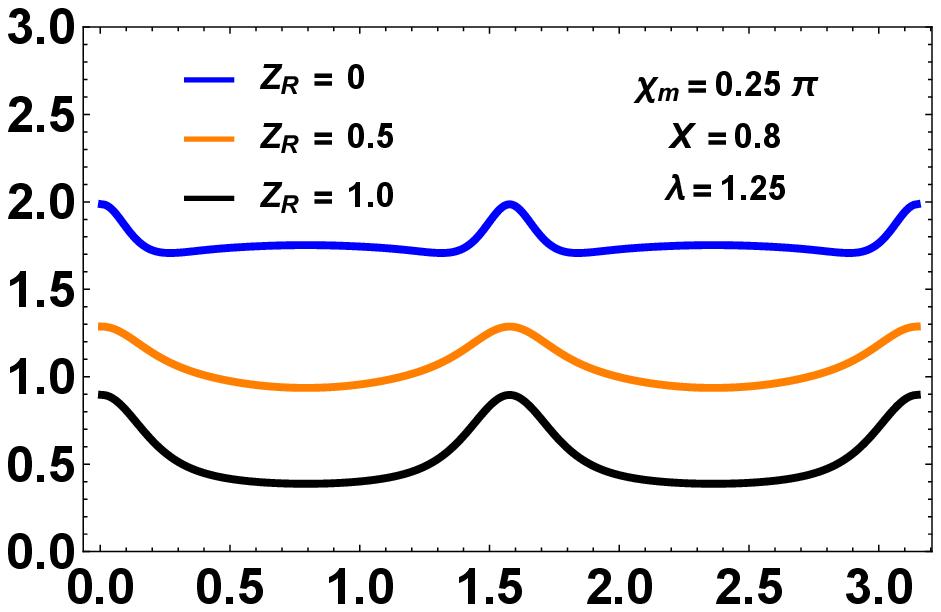}
\hspace{0.1cm}
\vspace{0.1cm}
\includegraphics[scale = 0.58]{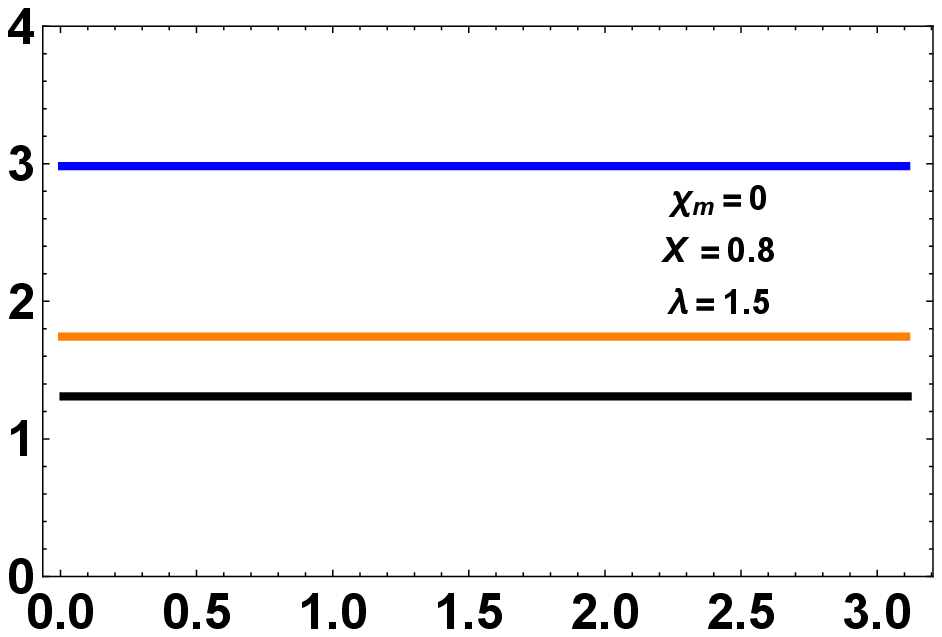}
\hspace{0.1cm}
\vspace{0.1cm}
\includegraphics[scale = 0.6]{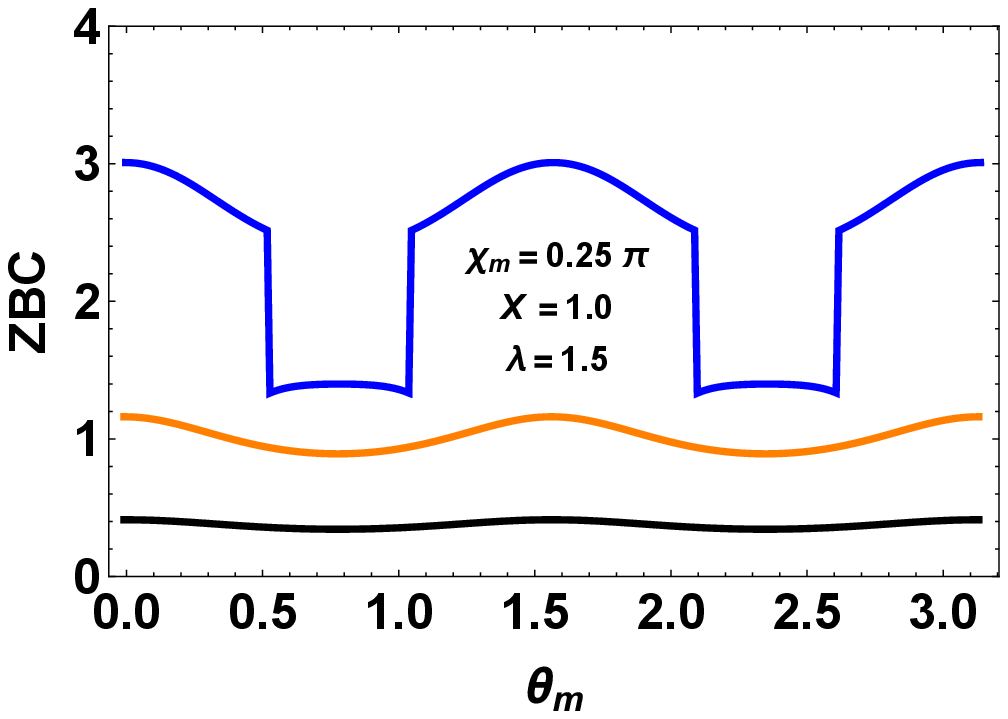}
\hspace{0.1cm}
\includegraphics[scale = 0.62]{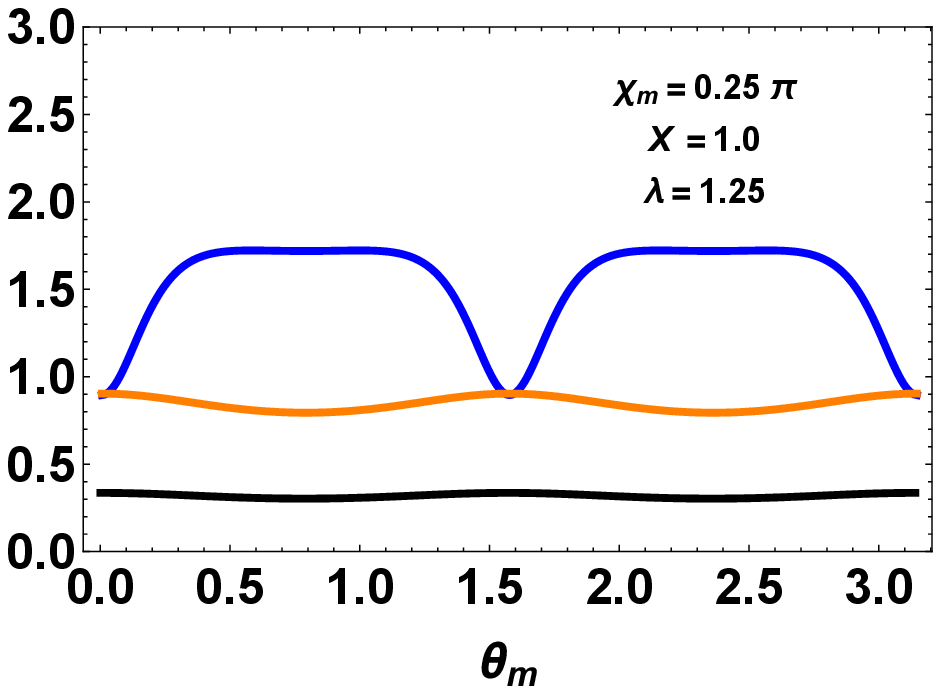}
\hspace{0.1cm}
\includegraphics[scale = 0.58]{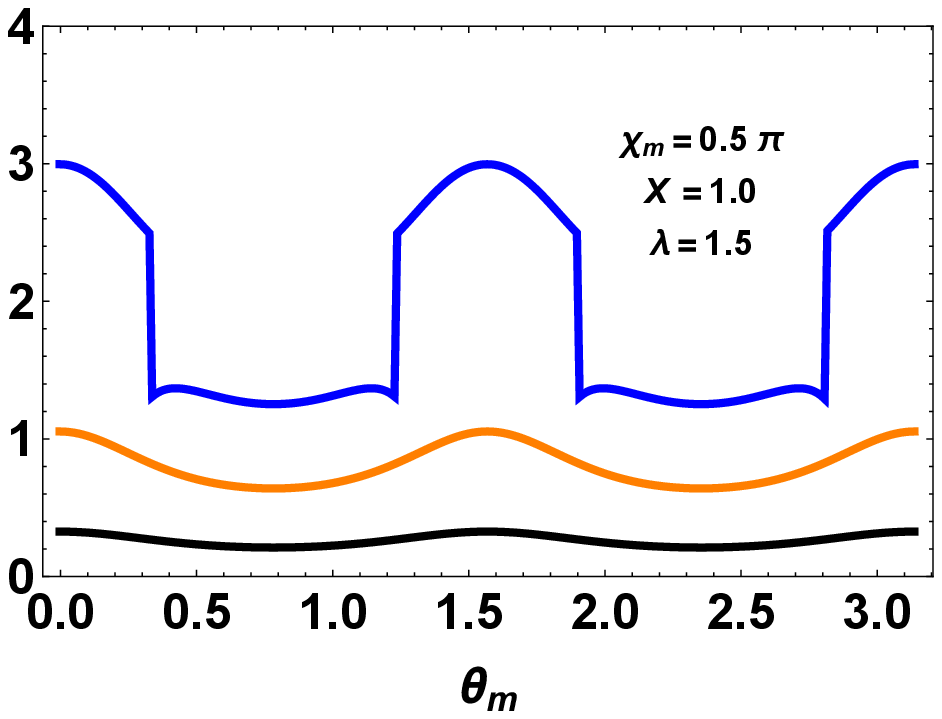}
\hspace{0.1cm}
}
\caption{Variation of ZBC with polar angle of manetization ($\theta_m$) for 
different magnetization strength $X$, azimuthal angle of magnetization 
$\chi_m$. The FWM are considered to be as $\lambda  = 1.25$ and $1.5$ 
respectively. The plot in the top left panel is for $X = 0.8$, while the
bottom left plot is for $X = 1.0$ with $\lambda = 1.5$. The middle line 
plots are drawn for $\lambda = 1.25$. The plot in the top right panel is for 
$\chi_m = 0$ and $X = 0.8$, while in the bottom right plot we consider 
$\chi_m = 0.5\pi$ and $X = 1.0$ keeping $\lambda = 1.5$ fixed.}
\label{fig6}
\end{figure*}

Due to the formation of Andreev bound states near $E = 
\Delta_-$ = $|\Delta_s - \frac{\Delta_t}{2}|$ and at  $\Delta_+$ = 
$|\Delta_s + \frac{\Delta_t}{2}|$, two sharp peaks in the conductance spectra 
are observed. The conductance is found to be maximum for Rashba free case, 
while it is significantly decreased in presence of RSOC as seen from all the 
plots of Fig.(\ref{fig2}). It is maximum for the low energy regions i.e. 
E$<\Delta_-$, then it gradually decreases for low Rashba cases i.e. for 
$Z_R = 0$ and $0.5$ while for
$Z_R = 1$, no significant change is seen. It shows a peak near $\Delta_-$ for 
$\theta_m = 0.25\pi$, $\chi_m = 0$ and $X = 0.8$  in all RSOC's. For the region 
$\Delta_-<E<\Delta_+$, in every cases it decreases rapidly and then rises monotonically
till $\Delta_+$. For all the plots a sharp peak it observed at $E = \Delta_+$. 
A similar characteristics is also seen for $X =1.0$. However,
in this case the decrease in conductance is quite gradual. For $\theta_m = 0.25\pi$, 
$\chi_m = 0.5\pi$ and $X = 1.0$, the conductance shows two very sharp peaks near 
$0.17\Delta_t$ and $0.83\Delta_t$ as shown in middle plot of bottom panel of 
Fig.(\ref{fig2}). 
A similar characteristics is also observed for $X = 0.8$ as seen from the 
left plot of Fig.(\ref{fig3}). 
The peak near $0.17\Delta_t$ nearly disappeared for $\theta_m = 0.5\pi$.
In Fig.(\ref{fig3}), we consider the effect of barrier strength $Z_0$. For this purpose, we consider
$\theta_m = 0.25\pi$, $\chi_m = 0.5\pi$, $X = 0.8$ and $\lambda = 1.5$. 
We studied the conductance characteristics for a transparent ($Z_0 = 0$), partially opaque
($Z_0 = 2$) and a strongly opaque ($Z_0 = 5$) barriers in Fig.(\ref{fig3}). The 
conductance has been analyzed for for Rashba free case  ($Z_R$ = 0), weak RSOC
($Z_R = 0.5$), and for moderate RSOC ($Z_R = 1$). Though the conductance gets 
decreased with the rise of RSOC as seen earlier, however a reverse trend is seen for $Z_0 = 2$.
In this case, it is found that the conductance gradually increases with the rise of RSOC. 
Another important point is that there is a suppression of broadening of the peak
which is quite significant for $Z_0 = 0$. For a strongly opaque barrier ($Z_0 = 5$), 
the effect of RSOC is not too significant as seen from the right plot of Fig(\ref{fig3}).
So, what is the role of barrier strength on RSOC and how it effects the maximum 
conductance?

In order to understand that it is necessary to investigate the variation of maximum 
normalized conductance G$_{max}$ with the RSOC ($Z_R$) for different barrier transparency ($Z_0$).
We have plotted the same in Fig.(\ref{fig4}) for  $Z_0 = 0, 2$ and $5$. The 
plot in the left is for $\lambda = 0.5$, while the right for $\lambda = 1.5$.  
It is seen that for a transparent barrier the maximum conductance G$_{max}$ shows 
a sharp peak near $Z_R \simeq 0.1$ and a gradual 
decrease is seen with the further increase in $Z_R$. Similar characteristics are also observed
for an opaque barrier. For $Z_0 = 2$, a sharp peak is seen at $Z_R \simeq 2$, which 
sharply decrease with the further rise of $Z_R$ as seen earlier for a transparent barrier. However,
for $Z_0  = 5$, it shows nearly a linear behaviour for low values of RSOC while rises sharply for higher 
values of $Z_R$. An exactly similar characteristics of $G_{max}$ is also observed for $\lambda = 1.5$. 
It is also observed that with the increase in FWM parameter $\lambda$, $G_{max}$ also increases.

We studied the effect of FWM on the conductance spectrum,
which is shown in Fig.(\ref{fig5}). For this purpose, we investigated the 
conductance spectra for three different choices of $\lambda$, viz., (i) for 
$q_{FS} < k_{FF}$ with $\lambda = 0.5$ (left plot), (ii) no mismatch 
$q_{FS} = k_{FF}$ with $\lambda = 1.0$ (middle 
plot), and (iii) for $q_{FS} > k_{FF}$ with $\lambda = 1.25$ (right plot). 
We consider, $\theta_m = 0.25\pi$, $\chi_m = 0.5\pi$, $X = 0.8$ and a  
transparent barrier with $Z_0 = 0$ for this analysis. It is seen that the 
characteristics of the conductance spectrum is independent of $\lambda$. 
However, rise of FWM parameter $\lambda$ enhances the 
conductance for all values of RSOC as seen from Fig.(\ref{fig5}). 

\subsubsection{Dependence of Zero Bias Conductance (ZBC) on the polar angle of magnetization $\theta_m$}
It should to be noted that there exist a peak even at $\theta_m = 0$ and
there exist a significant change with the polar angle of magnetization and 
magnetization strength as seen from Fig.(\ref{fig2}). Though there is a very 
small variation in conductance with the azimuthal angle, however its effect 
cannot be neglected. To understand the magnetization orientation dependence of 
conductance spectrum we have plotted the ZBC as a 
function of $\theta_m$ with different choices of $\chi_m$, $X$ and $\lambda$ in 
Fig.(\ref{fig6}). All the plots are drawn for a nearly transparent barrier 
($Z_0 = 0.1$) and for different RSOC coupling strengths $Z_R$. Though the 
effect of azimuthal angle is still need to be known, however for $\chi_m = 0$, 
the ZBC remains constant for all the choices of $X$ and $\lambda$. It is 
observed from all the plots that the ZBC plots are highly symmetrical about 
$\theta_m$ for non vanishing $\chi_m$. For $\chi_m = 0.25\pi$, $X = 0.8$ and 
$\lambda = 1.5$, it shows an oscillatory behaviour having a ZBC Peak (ZBCP) at 
$\theta_m = 0, 0.5\pi$ and at $\pi$ for all the RSOC's as seen from the left 
top plot of Fig.(\ref{fig6}). Usually the ZBCP appears in unconventional 
superconductors due to the different phases of the transmitted electronlike 
and holelike quasi particles in the superconducting region \cite{tanaka}. 
However, according to Zutic and Valls \cite{zutic1,zutic2}, the appearance of 
ZBCP is due to the FWM and it may also appear in conventional superconductors 
too. With the increase of $X$ to $1.0$, a significant change in the ZBC 
spectrum is observed for $Z_R = 0$. The ZBC characteristics remains quite 
similar as for $Z_R = 0.5$ and $1.0$ as seen from the plot in the left 
bottom of Fig.(\ref{fig6}). In both cases the only significant observation is 
that with the rise of magnetization strength, the ZBC drastically decreases 
with the increasing values of $Z_R$. The plots in the middle panel (vertical) 
are drawn for the FWM $\lambda = 1.25$ and for two different choices 
of magnetization strength respectively $0.8$ and $1.0$. For $\lambda = 1.25$ 
with $X = 0.8$, the conductance characteristics is exactly similar with 
$\lambda = 1.5$ for $Z_R = 0.5$ and $1.0$. The conductance characteristics 
shows a significant change for Rashba free case for $\lambda = 1.25$. In this 
condition the oscillatory characteristics nearly vanishes, it gradually falls 
from a maxima and then remains constant followed by a small peak at 
$\theta_m = 0.5\pi$ as seen from the middle plot of the top panel of 
Fig(\ref{fig6}). Another important result we have seen that with the increase 
of FWM, the conductance increases by a significant 
amount.  A totally opposite behaviour from $X = 0.8$ is observed for 
$X = 1.0$ with $\lambda = 1.25$. In this case, ZBC Dips (ZBCDs) are observed 
at $\theta_m =0, 0.5\pi$ and $\pi$, from which the ZBC gradually rises, 
saturates and falls to the next dip as seen from the middle plot of the bottom 
panel of Fig(\ref{fig6}).
It is to be noted that for non vanishing values of RSOC's, the ZBC 
characteristics are almost similar. So it can be concluded that the ZBC 
spectrum highly dependent on FWM parameter and the strength of magnetization. 
We have also investigated the ZBC for  $\chi_m = 0.5\pi$, $X = 1.0$ 
and $\lambda = 1.5$. In this condition a very small change in the behaviour of 
ZBC is observed for Rashba free case as obtained for $\chi_m = 0.25\pi$ (see
bottom left plot), however no significant change is observed with the rise of 
RSOC. So it can also be concluded that the ZBC spectrum is nearly independent 
for higher values of RSOC strength.

\begin{figure*}[hbt]
\centerline
\centerline{
\includegraphics[scale = 0.59]{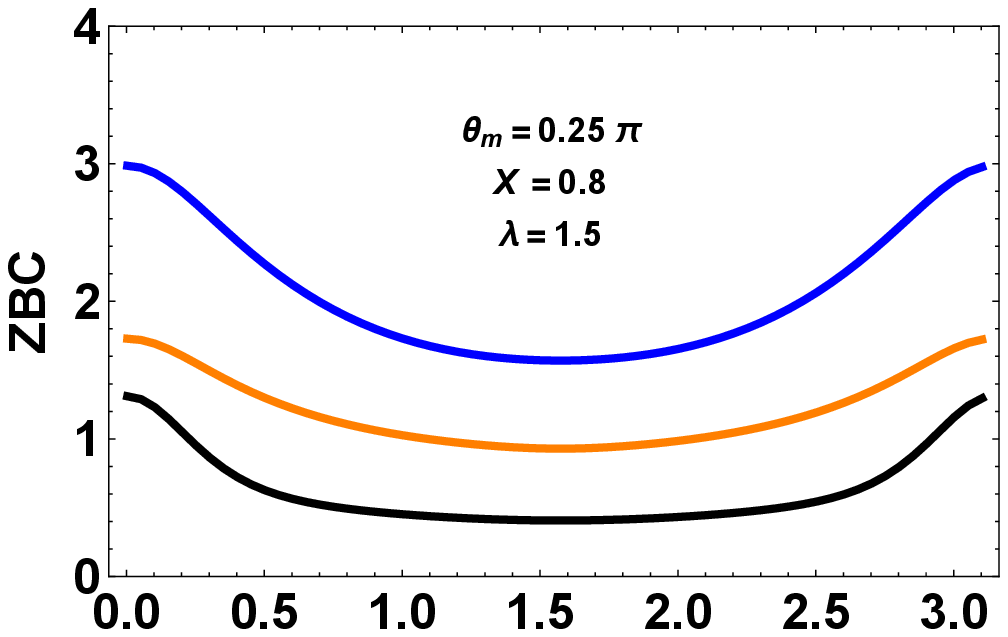}
\hspace{0.1cm}
\vspace{0.1cm}
\includegraphics[scale = 0.61]{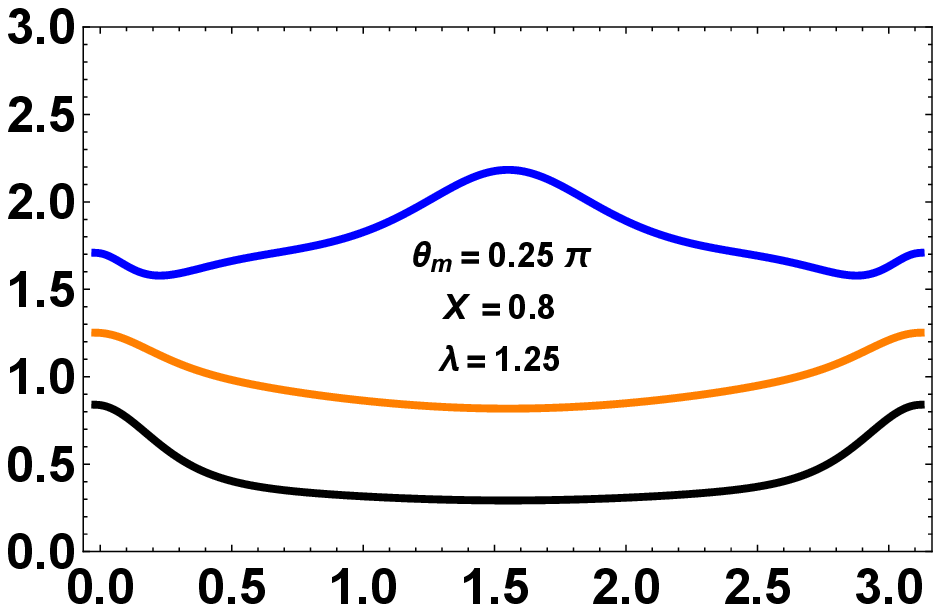}
\hspace{0.1cm}
\vspace{0.1cm}
\includegraphics[scale = 0.58]{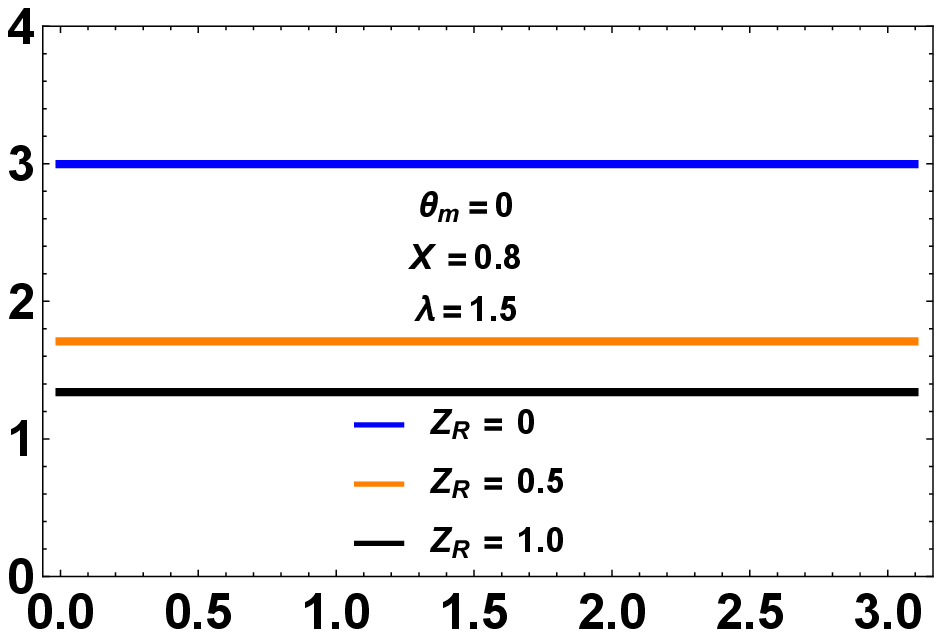}
\vspace{0.1cm}
\includegraphics[scale = 0.59]{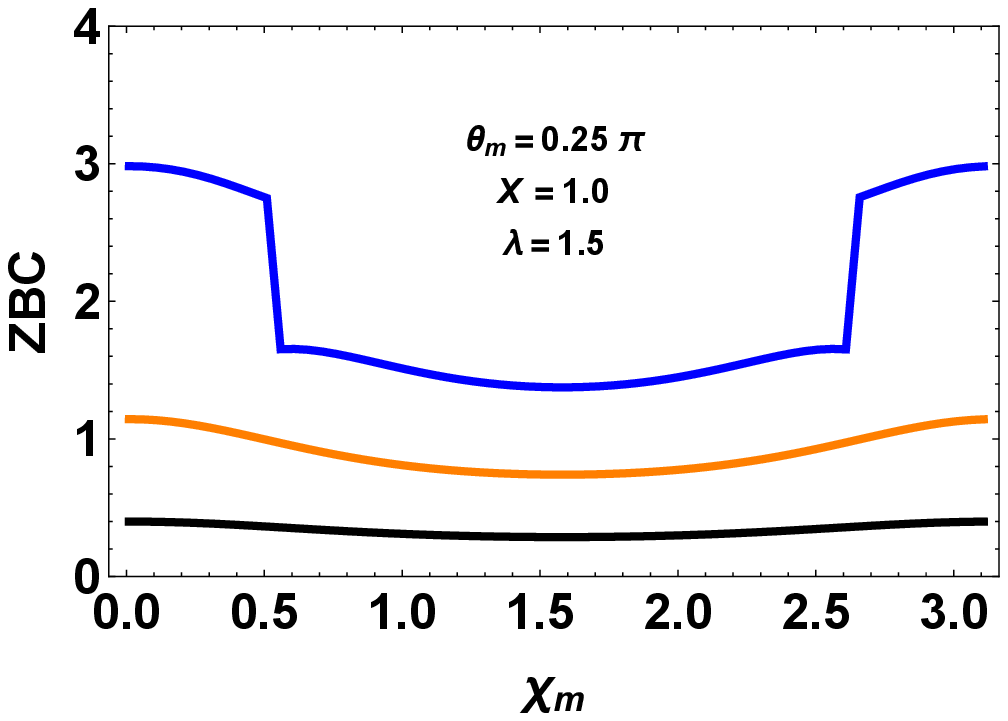}
\hspace{0.12cm}
\includegraphics[scale = 0.61]{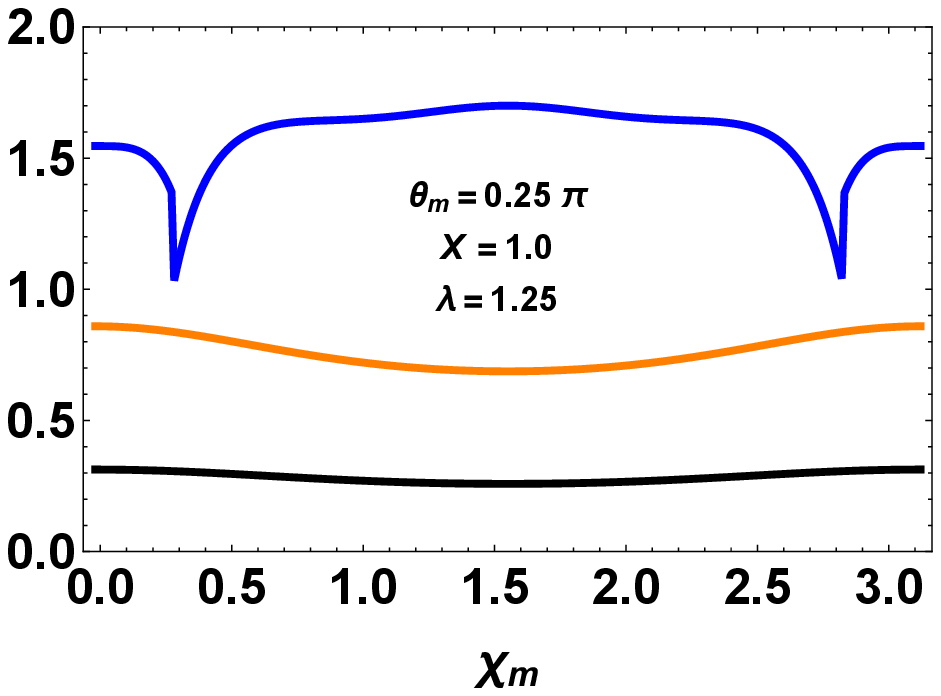}
\hspace{0.12cm}
\includegraphics[scale = 0.58]{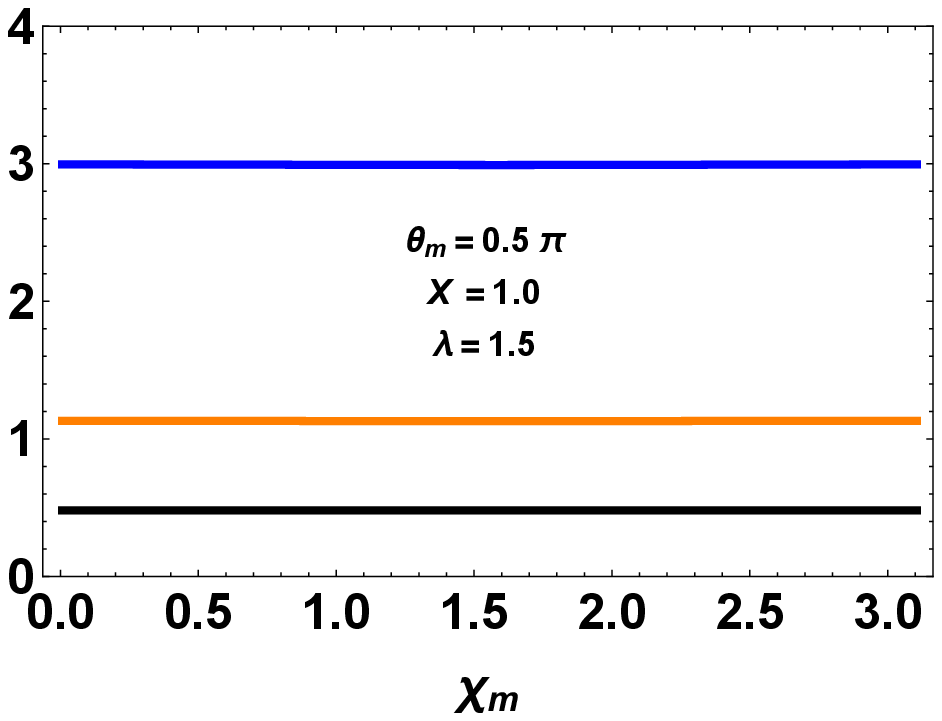}
}
\caption{Variation of ZBC with azimuthal angle of magnetization $\chi_m$ for 
different magnetization strength $X$ and polar angle of magnetization 
$\theta_m$. The FWM are considered to be as $\lambda = 1.25$ and $1.5$ 
respectively. The plot in the top left panel is for $X = 0.8$, while the 
bottom left panel plot is for $X = 1.0$ keeping $\lambda = 1.5$ fixed. The 
middle line plots are drawn for $\lambda = 1.25$. The plot in the top right 
panel is for $\theta_m = 0$ and $X = 0.8$, while in the bottom of
right plot we consider $\theta_m = 0.5\pi$ and $X = 1.0$ keeping 
$\lambda = 1.5$ fixed.}
\label{fig7}
\end{figure*}
\begin{figure*}[hbt]
\centerline
\centerline{
\includegraphics[scale = 0.58]{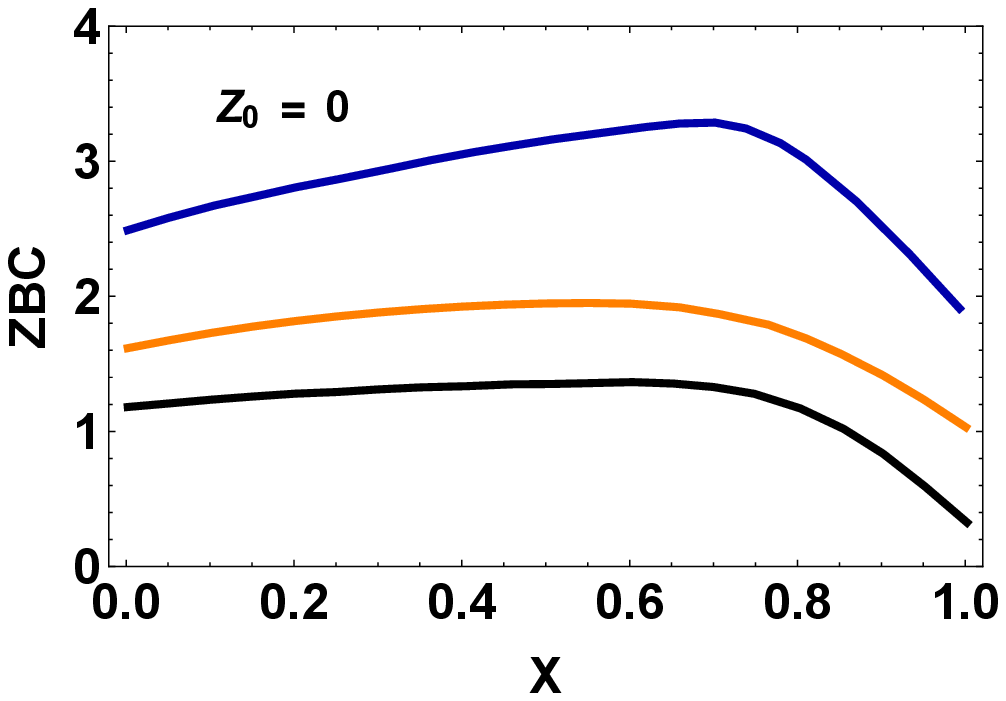}
\hspace{0.25cm}
\includegraphics[scale = 0.6]{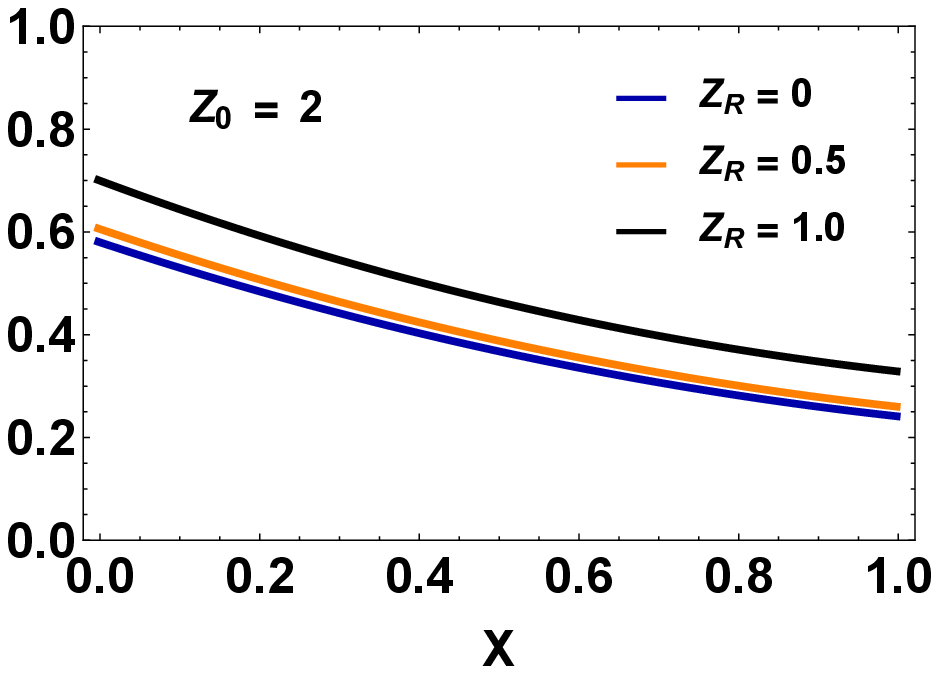}
\hspace{0.25cm}
\includegraphics[scale = 0.6]{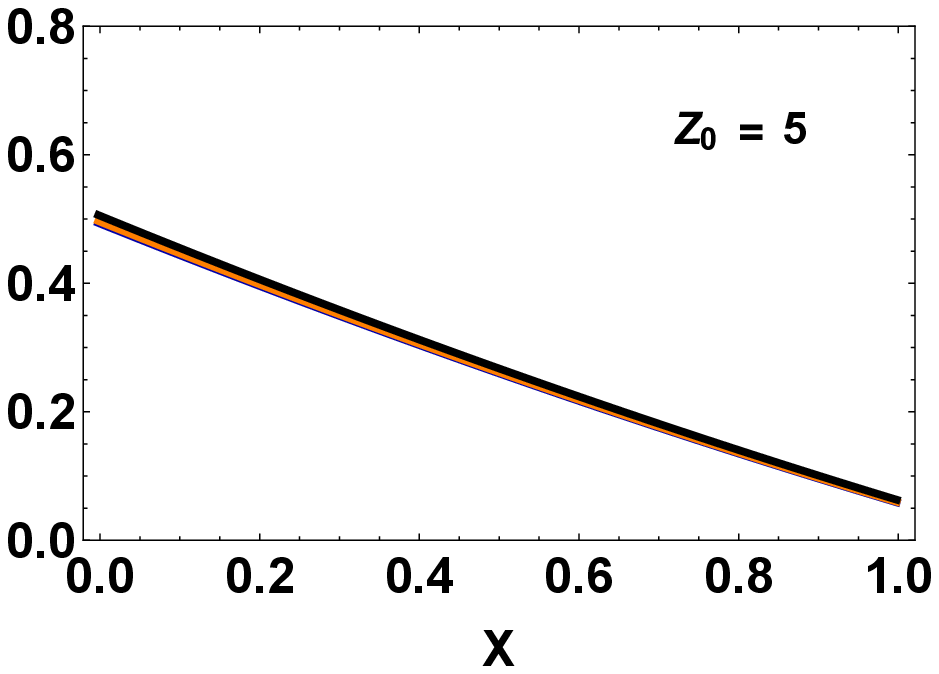}
\hspace{0.25cm}
}
\caption{Dependence of ZBC on the magnetization strength $X$ for different 
barrier strengths $Z_0$. The figures are plotted for $\lambda$ = $1.5$, 
$\theta_m = 0.25\pi$, $\chi_m = 0.5\pi$. The figure in the left is for 
$Z_0 = 0$,  middle figure is for $Z_0 = 2$, while the figure in the right is 
for $Z_0 = 5$.}
\label{fig8}
\end{figure*}

\subsubsection{Dependence of ZBC on the azimuthal angle of magnetization $\chi_m$}
The variation of ZBC with the azimuthal angle is shown in Fig.(\ref{fig7}) for
different FWM and the magnetization strength considering a transparent barrier
i.e. $Z_0 = 0.1$. We consider the polar angle of magnetization as $\theta_m = 0.25\pi$ for our analysis. However, we have also analyzed the ZBC characteristics 
for $\theta_m = 0$ and $0.5\pi$ as shown in the top and bottom of the right 
plots of the figure respectively. It is seen that the ZBC
behaviour is totally opposite for $\lambda = 1.25$ as observed from $1.5$. 
For $\lambda = 1.5$ a valley like pattern is seen while for $\lambda = 1.25$ a 
hill like characteristics is observed for Rashba free case. In the first case 
for $\lambda = 1.5$, the ZBC spectra has a minima at $\chi_m = 1.5\pi$, while 
for the preceding i.e. for $\lambda = 1.25$ it has a maxima at $\chi_m = 1.5\pi$ as seen from the middle top plot of Fig.(\ref{fig7}). It is to be noted 
that there is no significant change in ZBC with the change in the mismatch 
parameter $\lambda$ in presence of RSOC. The only observed change is 
with the rise in mismatch parameter $\lambda$ the conductance increases for 
all the values of RSOC as seen from the plots of Fig.(\ref{fig7}). It is also 
observed that for the polar angle $\theta_m = 0$ and $0.5\pi$, there exist no 
change in the ZBC characteristics. So it can be concluded from here that ZBC 
spectra is highly dependent on orientation of the polar angle too for Rashba 
free cases. For experimentally suitable conductance, arbitrary orientation of 
the polar angle is preferred. Furthermore, it can be concluded that with the 
rise of RSOC, the orientation dependence of ZBC is lost.

\subsubsection{Dependence of ZBC on the strength of magnetization $X$}

\begin{figure*}[hbt]
\centerline
\centerline{
\includegraphics[scale = 0.57]{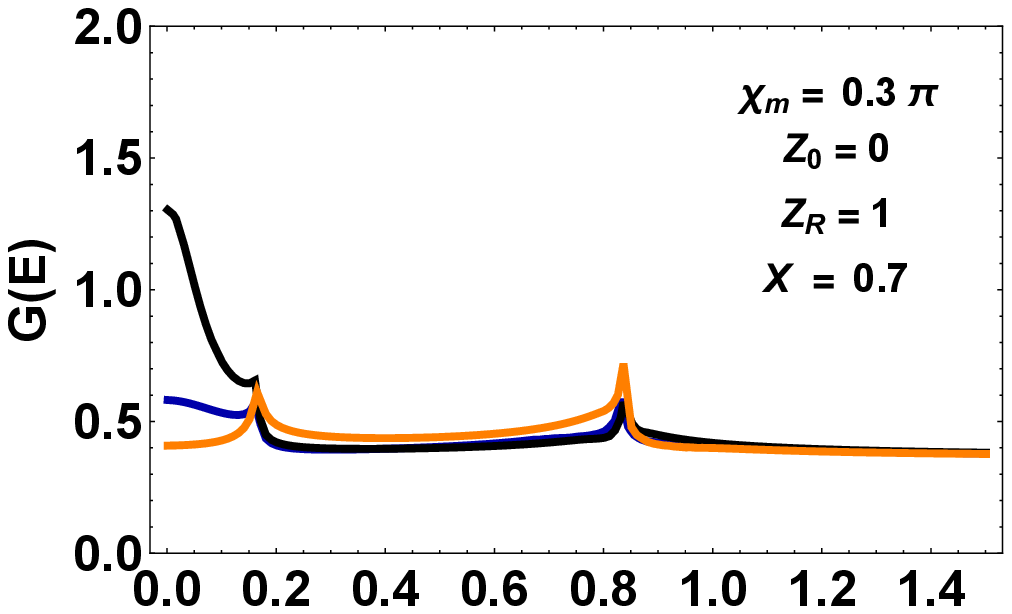}
\hspace{0.25cm}
\vspace{0.2cm}
\includegraphics[scale = 0.57]{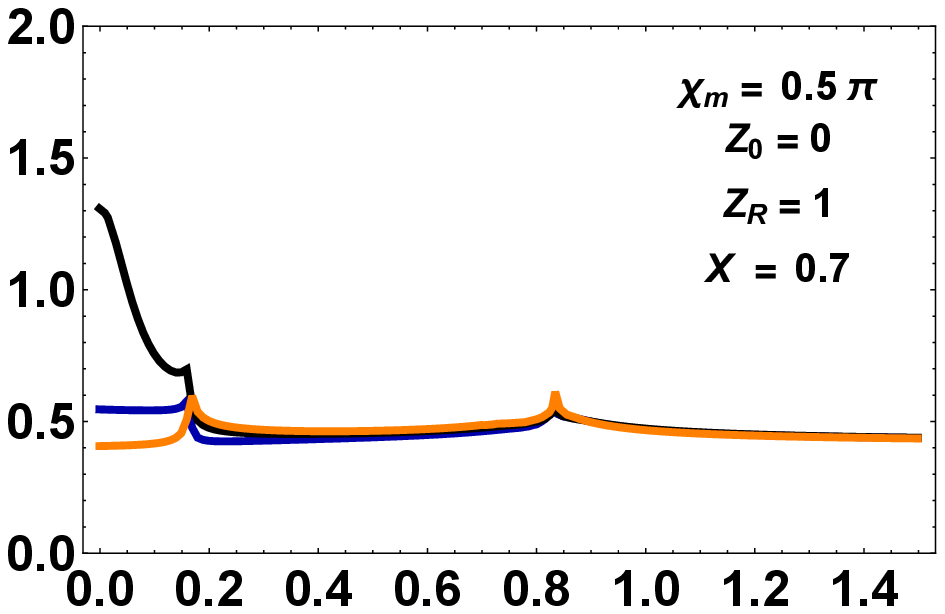}
\hspace{0.25cm}
\vspace{0.2cm}
\includegraphics[scale = 0.57]{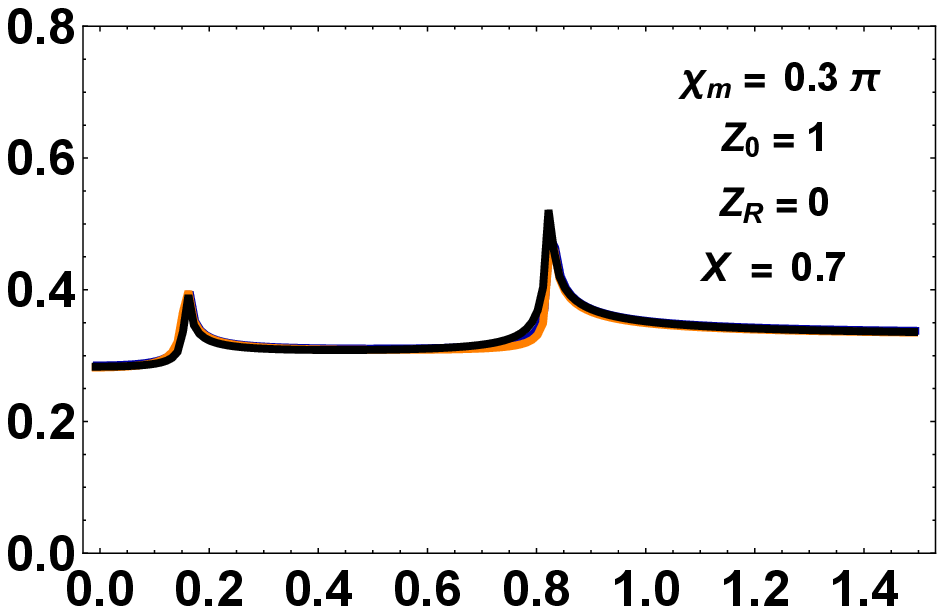}
\hspace{0.25cm}
\vspace{0.2cm}
\includegraphics[scale = 0.58]{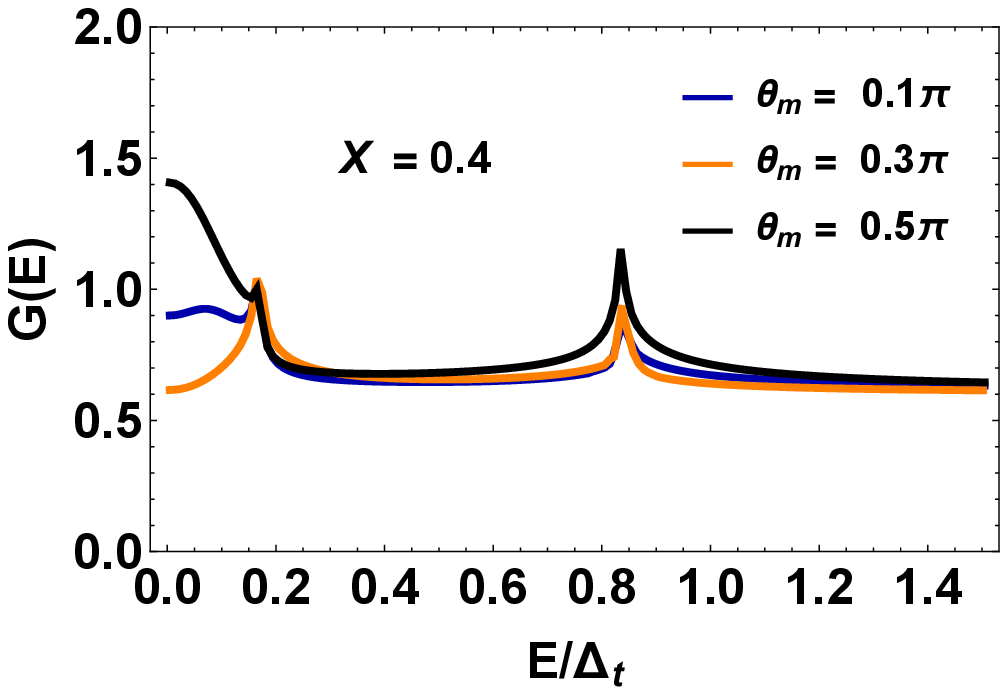}
\hspace{0.25cm}
\includegraphics[scale = 0.58]{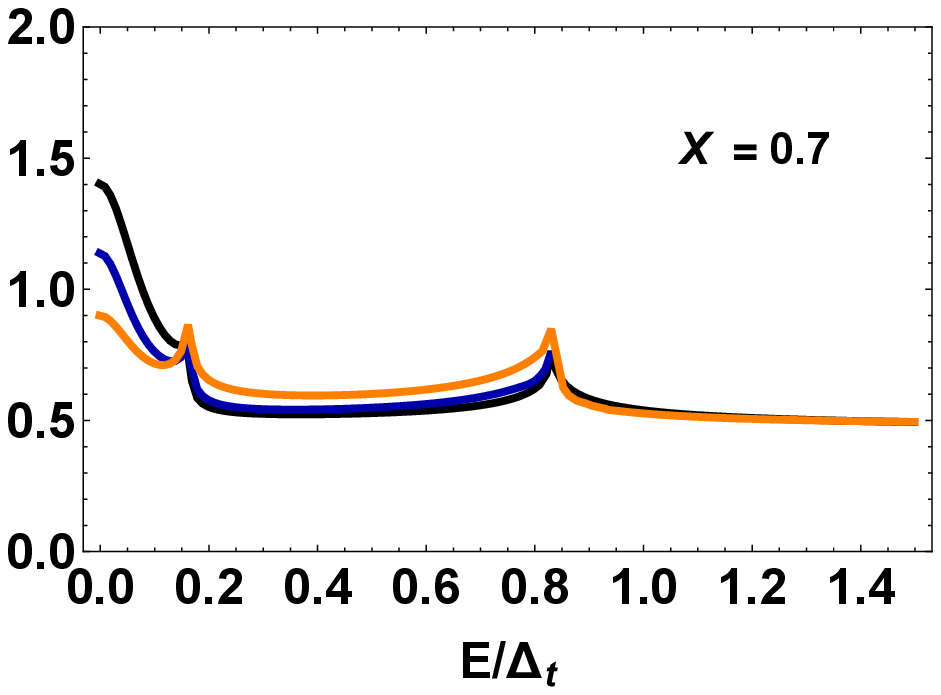}
\hspace{0.25cm}
\includegraphics[scale = 0.58]{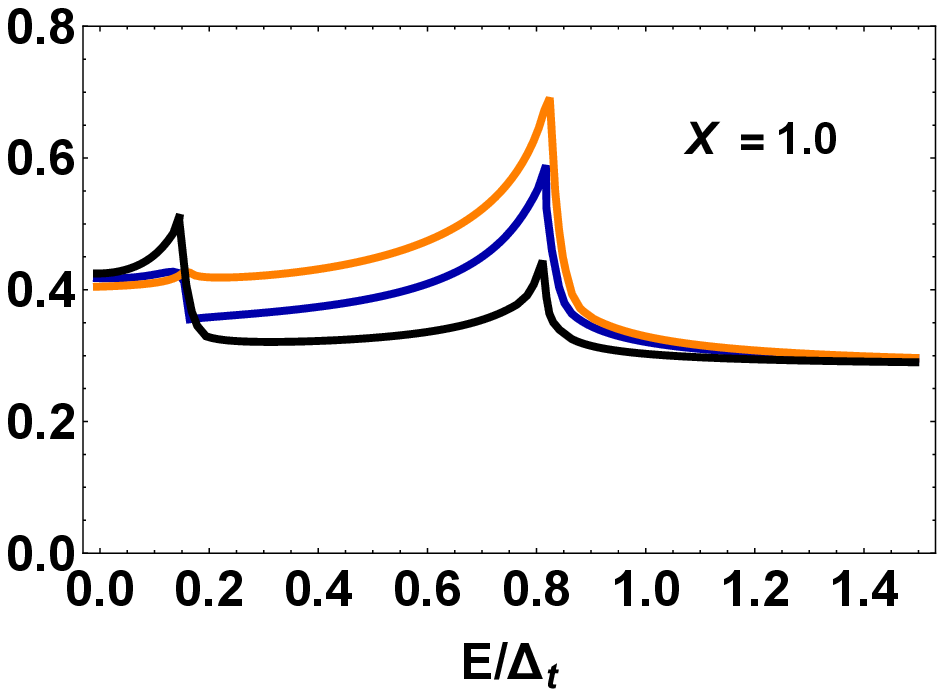}
\hspace{0.25cm}
}
\caption{Conductance spectra for different the azimuthal angle of magnetization 
$\chi_m$ and the strength of magnetization $X$. The left and the middle 
figures in the top panel are plotted respectively for $\chi_m = 0.3\pi$ and 
$0.5\pi$ with $Z_0 = 0$ and $Z_R = 1$. The figure in the top right is plotted 
for  $\chi_m = 0.3\pi$, $Z_0 = 1$ and $Z_R = 0$. The bottom figure in the left 
is for $X = 0.4$, middle is for $X = 0.7$ and the right is for $X = 1.0$. For 
the bottom figures we choose $\chi_m = 0.1\pi$,  $Z_0 = 0$ and $Z_R = 1$.}
\label{fig9}
\end{figure*}

To understand the interplay of magnetization with RSOC and their role on 
conductance characteristics, we studied the variation of ZBC spectra with 
the magnetization strength ($X$) in Fig.(\ref{fig8}). For this purpose we 
choose three different types of barriers mentioned in the earlier occasions,
viz., $Z_0 =  0, 2$ and $5$ respectively for a highly transparent, 
partially opaque and fully opaque barriers. Moreover, for 
this analysis we choose $\theta_m = 0.25\pi$, $\chi_m$ = $0.5 \pi$ and 
$\lambda = 1.5$. It is seen that for a highly transparent barrier the ZBC 
spectra shows a gradual rise with the rise in $X$. A ZBCP is observed at 
$X \approx 0.7$ for all Rashba cases. However, further rise of magnetization 
strength leads to decrease in the conductance as seen from the left plot of 
Fig.(\ref{fig8}). The ZBC characteristics drastically changes for a partially 
opaque and a strongly opaque barrier. In both the cases it is seen that the 
conductance decreases almost linearly with the rise of strength of the 
magnetization for all Rashba cases as seen from the middle and the right plots
of the figure.
Another significant result is that for a partially opaque and fully opaque 
barrier i.e. $Z_0 = 2$ and $5$ all Rashba cases provide maximum conductance at 
$X \approx 0$. The characteristics of the ZBC spectra are 
found to be nearly similar for all Rashba cases for opaque barriers. 
It is observed that for a transparent barrier, Rashba free cases provide the 
maximum 
conductance and it decreases with the increasing values of RSOC, while for a 
partially opaque barrier an opposite behaviour is observed. In this case with 
the rise of RSOC, the conductance increases. It is to be noted that for a 
strongly opaque barrier i.e. $Z_0 = 5$, the effect of RSOC on the ZBC is almost 
insignificant. So, it 
can be concluded here that a highly transparent barrier with moderate RSOC along with moderate 
magnetization strength is highly suitable for an experimentally realistic scenario.

\subsection{Effect of Magnetization}
\begin{figure*}[hbt]
\centerline
\centerline{
\includegraphics[scale = 0.57]{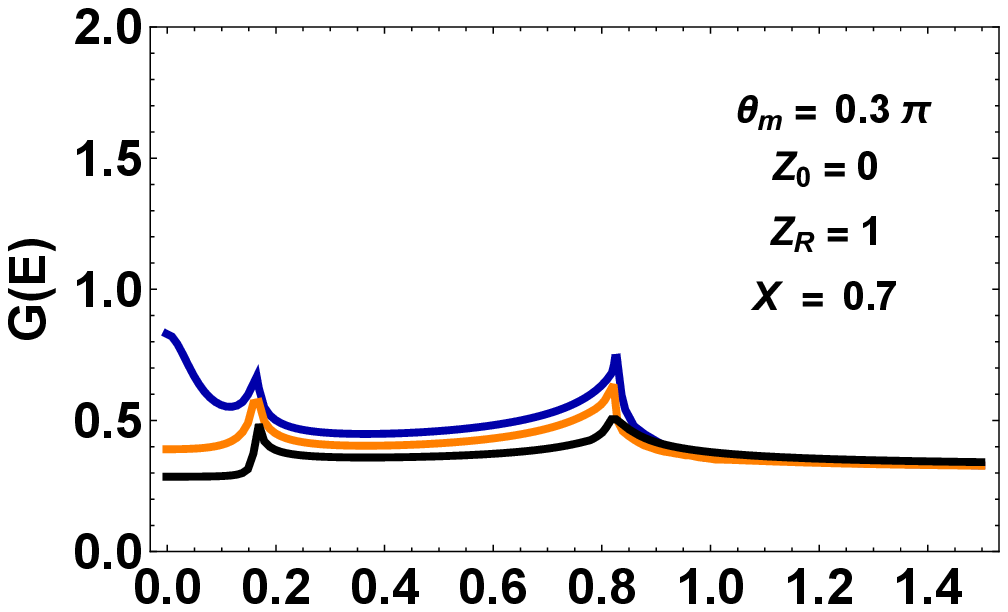}
\hspace{0.25cm}
\vspace{0.2cm}
\includegraphics[scale = 0.57]{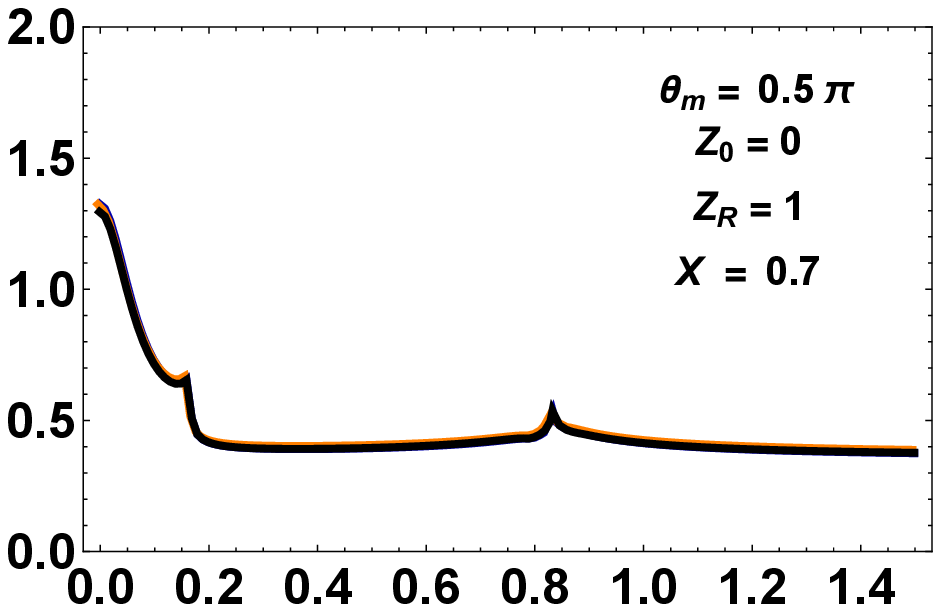}
\hspace{0.25cm}
\vspace{0.2cm}
\includegraphics[scale = 0.57]{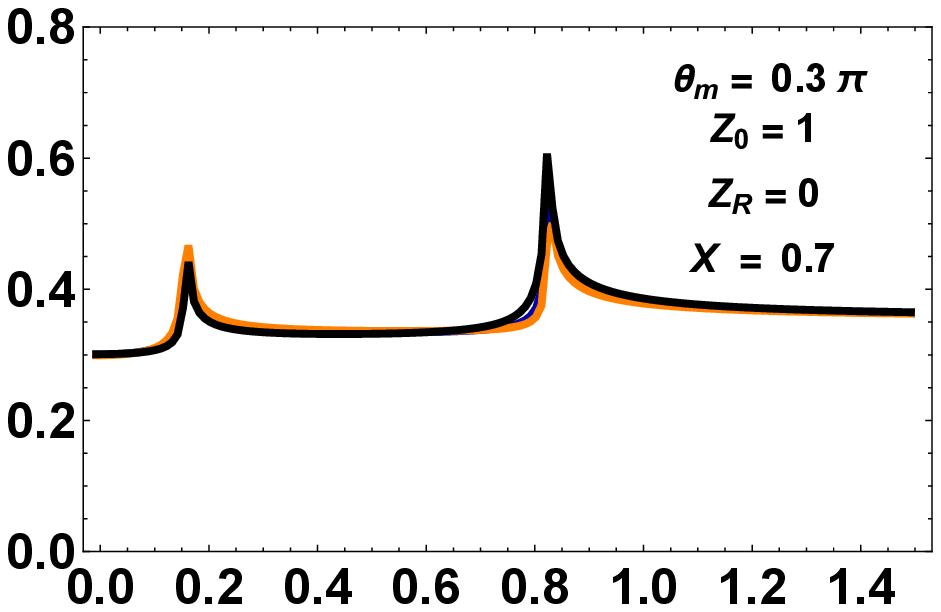}
\hspace{0.25cm}
\vspace{0.2cm}
\includegraphics[scale = 0.58]{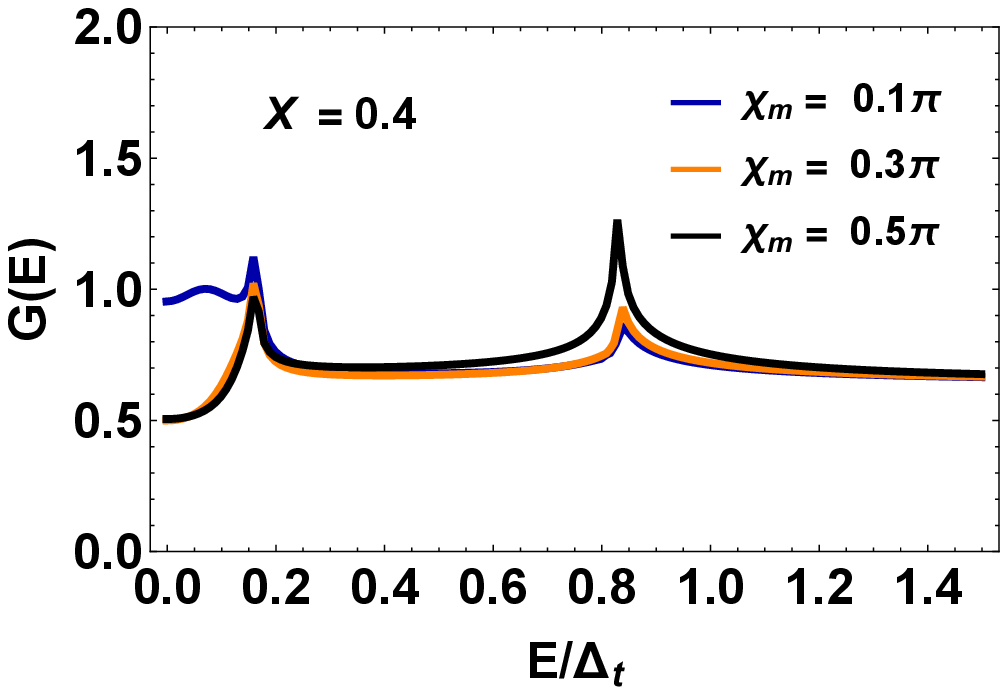}
\hspace{0.25cm}
\includegraphics[scale = 0.58]{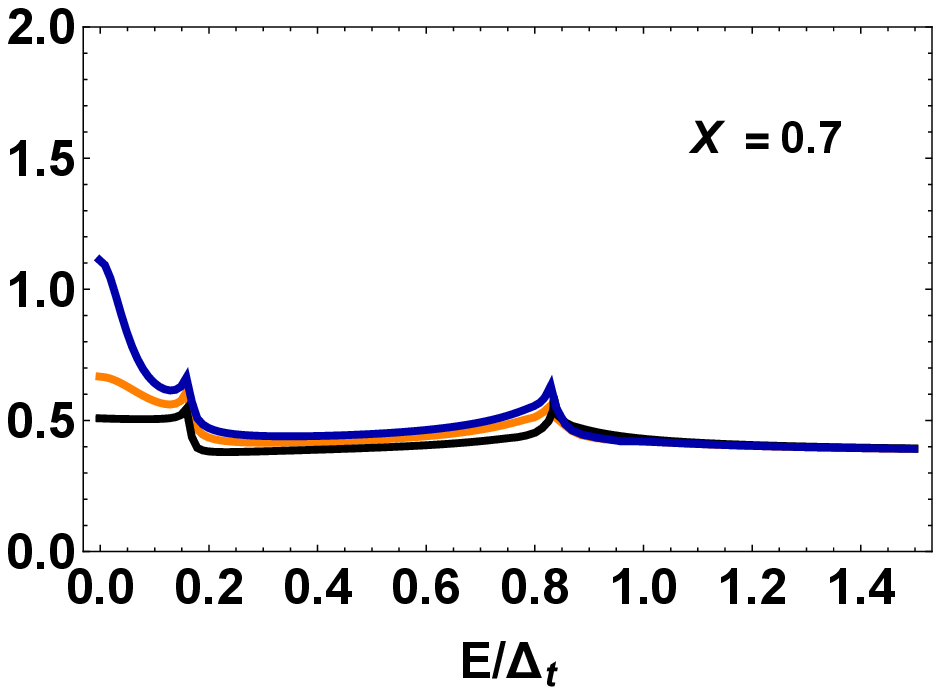}
\hspace{0.25cm}
\includegraphics[scale = 0.58]{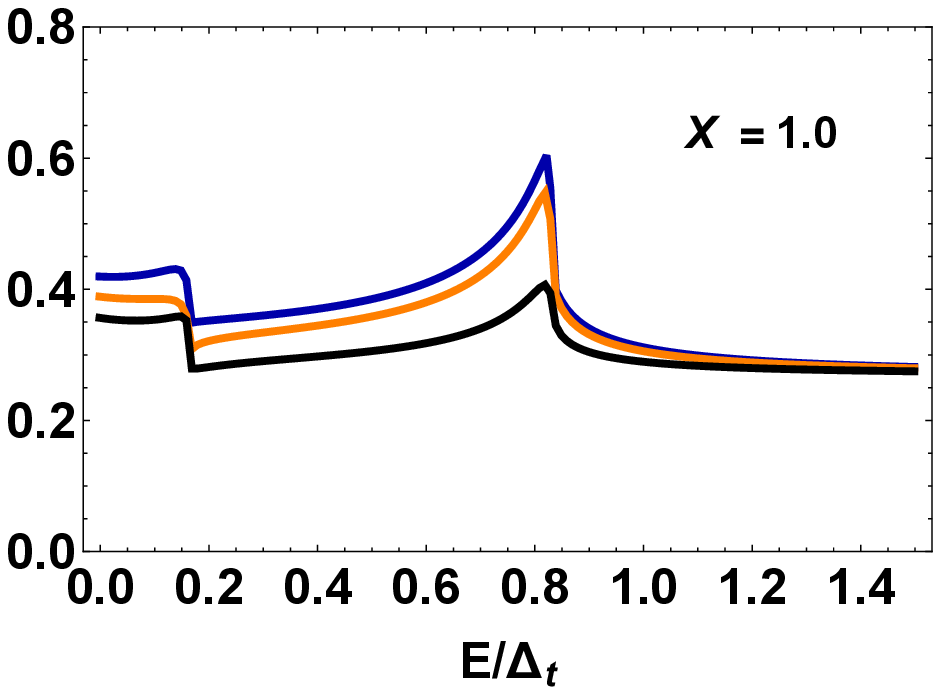}
\hspace{0.25cm}
}
\caption{Conductance spectra for different the polar angle of magnetization 
$\theta_m$ and the strength of magnetization $X$. The left and the middle 
figures in the top panel are plotted respectively for $\theta_m = 0.3\pi$ and 
$0.5\pi$ with $Z_0 = 0$ and $Z_R = 1$. The figure in the top right is plotted 
for $\theta_m = 0.3\pi$, $Z_0 = 1$ and $Z_R = 0$. The bottom figure in the left is for $X = 0.4$, middle is for $X = 0.7$ and the right is for $X = 1.0$. For 
the bottom figures we choose $\theta_m = 0.1\pi$,  $Z_0 = 0$ and $Z_R = 1$}
\label{fig10}
\end{figure*}

It is clear from Figs.(\ref{fig6}), (\ref{fig7}) and (\ref{fig8}) of the 
preceding section that
the ZBC spectra is highly dependent on magnetization. 
It is seen that conductance not only depend on the strength of magnetization 
but also dependent on the orientation of magnetization. To understand the role of magnetization more clearly and its interplay with RSOC,
we have plotted the conductance $G(E)$ as appear in the Eq.(\ref{eq21}) with 
the biasing energy $E$ scaled by 
the gap parameter $\Delta_{\pm}$ for different choices of the strength of magnetization $X$. For this purpose, we initially consider a highly transparent 
barrier i.e. $Z_0 = 0$. We have also 
investigated the same for a partially opaque barrier i.e. $Z_0 = 1$. We
vary the polar angle of magnetization $\theta_m$ keeping the azimuthal angle 
$\chi_m$ and the strength of magnetization $X$ fixed. The results are shown in 
the plots appear on the top panel of Fig.(\ref{fig9}). We have also analyzed 
the conductance characteristics by varying azimuthal angle of magnetization 
$\chi_m$ keeping the polar angle $\theta_m$ and the strength of magnetization 
$X$ fixed. This results are shown in plots on the top panel of 
Fig.(\ref{fig10}). The plots on the bottom panel of Figs.(\ref{fig9}) and 
(\ref{fig10}) represent the conductance spectra for different strength of $X$ 
keeping $\chi_m$ and $\theta_m$ constant respectively.

\begin{figure*}[hbt]
\centerline
\centerline{
\includegraphics[scale = 0.57]{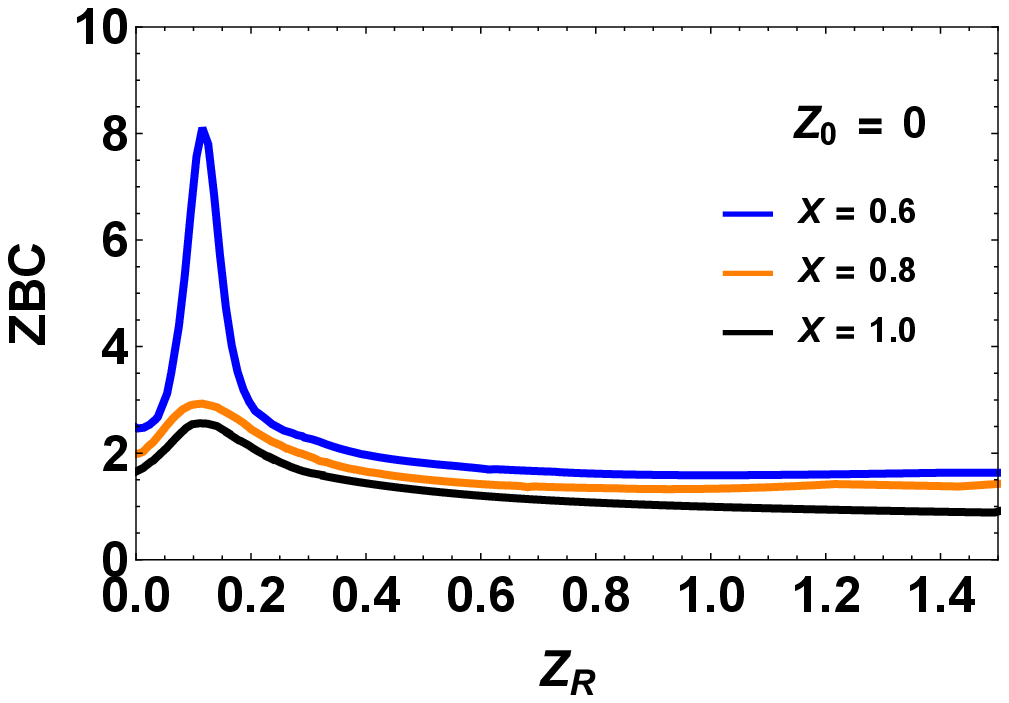}
\hspace{0.25cm}
\includegraphics[scale = 0.58]{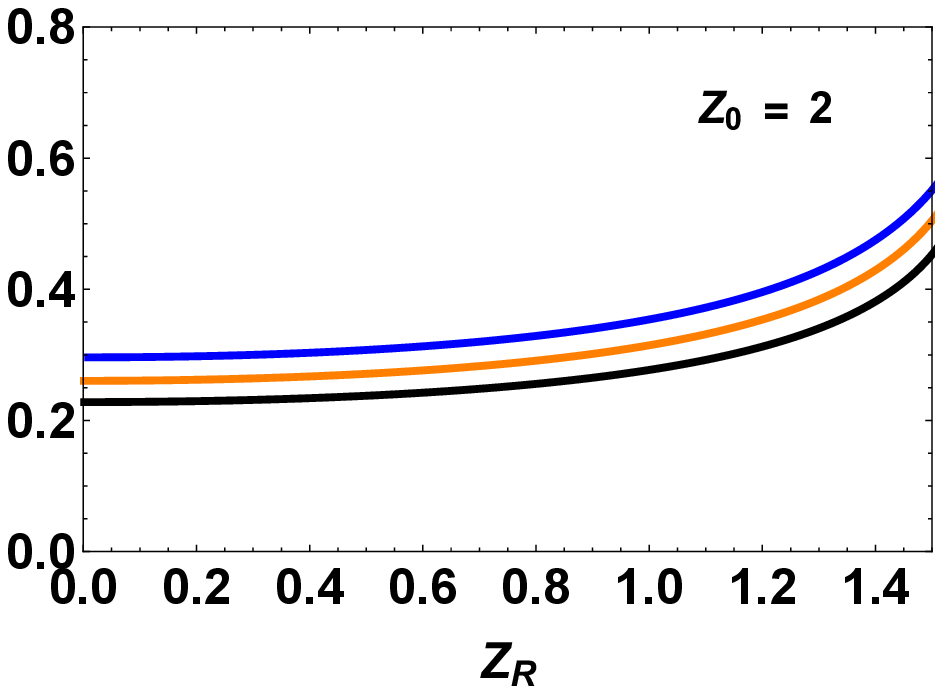}
\hspace{0.25cm}
\includegraphics[scale = 0.59]{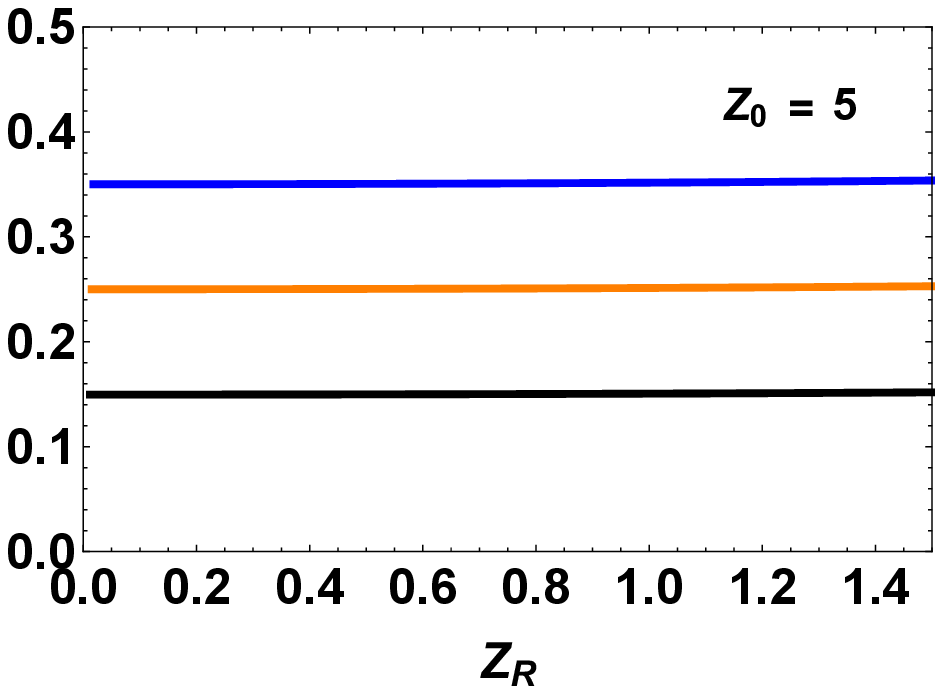}
\hspace{0.25cm}
}
\caption{Dependence of ZBC on the RSOC ($Z_R$) for different barrier strengths 
$Z_0$. The figures are plotted for $\lambda = 1.5$, $\theta_m = 0.5\pi$, 
$\chi_m = 0.5\pi$. The figure in the left is for $Z_0 = 0$, middle figure is 
for $Z_0 = 2$, while the figure in the right is for $Z_0 = 5$.}
\label{fig11}
\end{figure*}

To understand the significance of the azimuthal angle $\chi_m$ we plotted the 
conductance spectra considering $Z_0 = 0$, $Z_R = 1$ and $X = 0.7$ for 
$\chi_m = 0.1\pi$ (middle of bottom panel), $\chi_m = 0.3\pi$ (left of top 
panel) and $\chi_m = 0.5\pi$ (middle of top panel) in Fig.(\ref{fig9}). For all our analysis here we consider FWM $\lambda = 1.5$. It is seen that the pattern 
of conductance spectrum is quite similar for all studied orientations of 
$\chi_m$. The conductance is found to be maximum for $\chi_m = 0.1\pi$ with 
$X = 0.7$. In this case a sharp peak is observed as $E$ approaches to 
$\Delta_-$ and $\Delta_+$ similar to the spectra seen earlier for different 
RSOC's. Another important point is that the conductance characteristics 
is nearly similar for arbitrary ($\theta_m = 0.3\pi$) and perpendicular 
orientations ($\theta_m = 0.5\pi$) of the polar angle of magnetization 
$\theta_m$ with $\chi_m = 0.5\pi$ and $X = 0.7$. 
For perpendicular orientation of $\theta_m$, the sharpness of the conductance 
peak is found to be minimum as seen from the plot on the middle of top panel. 
we are also interested to see the behaviour of conductance spectrum for a 
partially opaque barrier with $Z_0 = 1$. It is observed that for an opaque 
barrier the conductance spectrum is totally independent on the strength as 
well as the orientation of polar angle of magnetization as seen from the 
plot at the right of the top panel of Fig.(\ref{fig9}). We have also studied 
the conductance spectra for three different choices of magnetization strength, 
viz., $X = 0.4$ (left of bottom panel), $X = 0.7$ (middle of bottom panel) 
and $X = 1.0$ (right of bottom panel) with $\chi_m = 0.1\pi$, $Z_0 = 0$ and 
$Z_R = 1$. It is to be noted here that for $M<<E_F$, the conductance is found 
to be maximum for $\theta_m = 0.5\pi$, while it is found to be minimum 
as $M\rightarrow E_F$. In this situation, the arbitrary configuration of 
$\theta_m$ shows maximum conductance as seen from the plots of middle and 
right of the bottom panel of Fig.(\ref{fig9}). It is also observed that the 
conductance spectrum gets suppressed with the rise of magnetization strength. 

It is also our interest to see the interplay of polar angle of magnetization 
$\theta_m$ with conductance. Thus we studied the conductance spectra 
considering $Z_0 = 0$, $Z_R = 1$ and $X = 0.7$ for $\theta_m = 0.1\pi$ (middle 
of bottom panel), $\theta_m = 0.3\pi$ (left of top panel) and 
$\theta_m = 0.5\pi$ (middle of top panel) as shown in Fig.(\ref{fig10}). It is
seen that for perpendicular orientation of polar angle of magnetization, the 
conductance spectra are totally independent on the orientation of azimuthal 
angle $\chi_m$ of magnetization. Though the two peaks appear in all three 
situations but the sharpness of the peak is very small in this condition. 
For $\theta_m = 0.1\pi$ and $0.3\pi$, the conductance is found to be maximum 
for $\chi_m = 0.1\pi$, while it is minimum for $\chi_m = 0.5\pi$ as seen from 
the plots of Fig.(\ref{fig10}) as mentioned above. It is also noted that with 
the 
decrease in the transparency of the barrier with $Z_0 = 1$, the conductance 
characteristics is almost independent on the orientation of azimuthal angle of 
magnetization as seen from the plot on top right panel of Fig.(\ref{fig10}) 
as seen earlier in Fig.(\ref{fig9}). We have also studied the conductance 
spectra for three different choices of magnetization strength viz., $X = 0.4$ 
(left of bottom panel), $X = 0.7$ (middle of bottom panel) and $X = 1.0$ (right of bottom panel) with $\theta_m = 0.1\pi$, $Z_0 = 0$ and $Z_R = 1$. Another 
important point is for $M<<E_F$, the conductance is found to be maximum for 
$\chi_m = 0.5\pi$, while it is found to be minimum as $M\rightarrow E_F$ as in
the case for $\theta_m = 0.5\pi$. In
this situation, the orientation $\chi_m = 0.1\pi$ shows maximum conductance as 
shown in the plots of middle and right of the bottom panel of Fig.(\ref{fig10}).

We are also interested to see the variation of ZBC spectra for different 
strength of magnetization and the barrier transparency. For this purpose we 
studied the variation of ZBC spectra with RSOC parameter $Z_R$ for a 
transparent barrier ($Z_0 = 0$), partially opaque barrier ($Z_0 = 2$) and a 
strongly opaque barrier ($Z_0 = 5$), which is shown in Fig.(\ref{fig11}). In 
all the cases, it is seen that with the rise in the strength of magnetization 
$X$ the ZBC decreases. Also, with the decrease in transparency of the barrier 
the conductance decreases. The ZBC is found to be maximum for $X = 0.6$, while 
it is decreases as $M\rightarrow E_F$. A sharp ZBCP is seen for transparent 
barrier nearly at $Z_R = 0.1$ for all choices of $X$. However, the sharpness 
of the peak is found to be maximum for $X = 0.6$. With the increase in RSOC, 
the ZBC decreases monotonically for $Z_0 = 0.$ A similar characteristics in 
ZBC spectra is also seen for a partially opaque barrier with $Z_0 = 2$.
However, in this case the ZBC increases with $Z_R$ and hence the ZBCP appears 
for the higher values of $Z_R$ as seen from the middle plot of 
Fig.(\ref{fig11}). So, it can be concluded that the with the decrease in 
transparency of the barrier the ZBCP appear for higher values of RSOC. For a 
strongly opaque barrier ZBC characteristics is found to be totally independent 
of RSOC. 

\begin{figure}[hbt]
\centerline
\centerline{
\includegraphics[scale = 0.75]{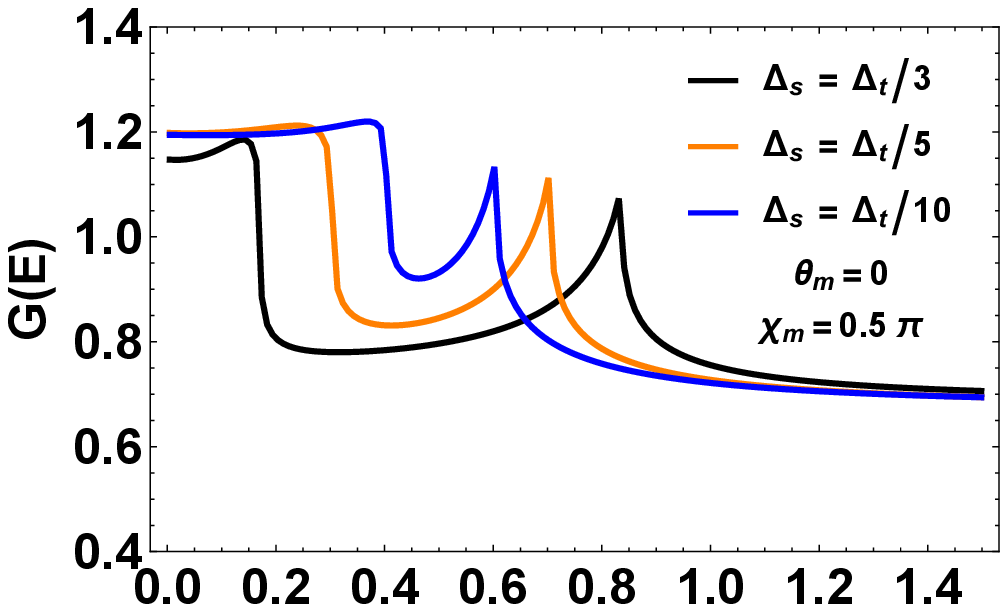}
\vspace{0.2cm}
\includegraphics[scale = 0.75]{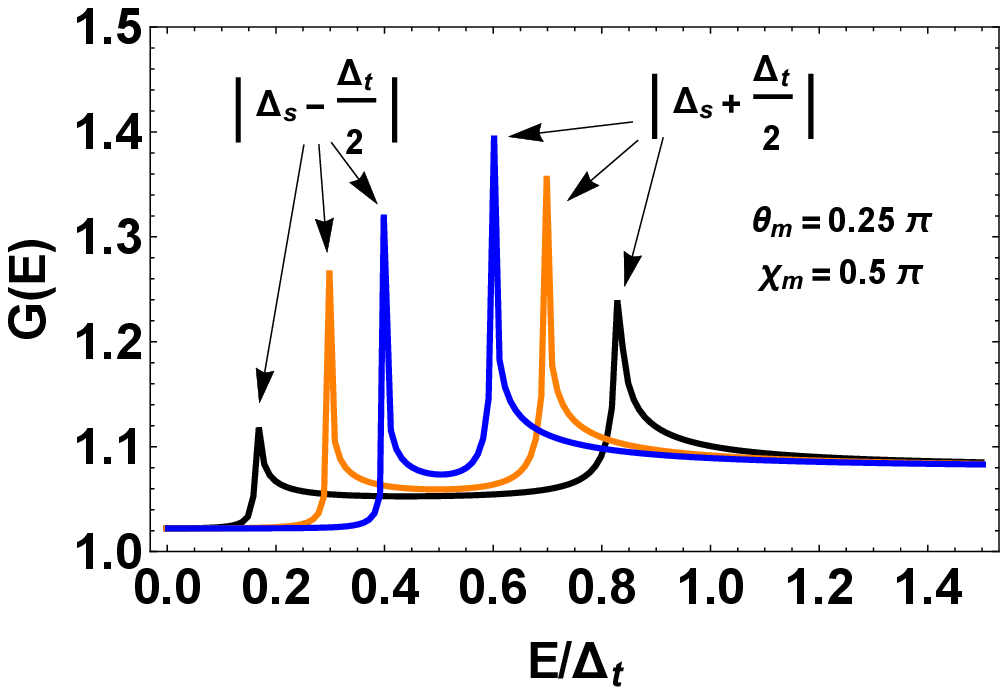}
\vspace{0.2cm}
}
\caption{Conductance spectra for different mixing of singlet-triplet gap 
parameter with magnetization strength $X = 1.0$, azimuthal angle of 
magnetization $\chi_m = 0.5\pi$, RSOC parameter $Z_R = 0.5$ and barrier 
strength $Z_0 = 0$. The plot in the top panel is for polar angle
$\theta_m = 0$, while the plot in the bottom panel is for 
$\theta_m = 0.25\pi$. The three different lines represent different choices of 
$\Delta_s$.}
\label{fig12}
\end{figure}   

\subsection{Effect of different singlet-triplet mixing ratio}
It is of our interest to see what happens to the conductance for different 
spin singlet-triplet mixing. Also it was found that a highly transparent 
barrier is often realizable from many Scanning Tunneling Microscope (STM) 
experiments \cite{linder,iniotakis}. So in view of the experimentally suitable 
situation we studied the conductance spectrum of a highly transparent barrier 
with $Z_0 = 0$ for three different spin singlet-triplet mixing as shown in 
Fig.(\ref{fig12}). We consider the RSOC parameter $Z_R = 0.5$, FWM parameter 
$\lambda = 1.5$ , azimuthal angle of magnetization $\chi_m = 0.5\pi$ and the 
strength of magnetization $X = 1.0$ for this analysis. For the figure on the 
top we consider the polar angle of magnetization $\theta_m = 0$, while for the 
bottom figure it is $0.5\pi$. It is seen that with the increase in the ratio 
of $\Delta_t:\Delta_s$, the appearance of the conductance peaks at 
$\Delta_-$ and $\Delta_+$ is quite nearer each other. It is also to be noted that
with the increase in the ratio of $\Delta_t:\Delta_s$, the sharpness of the 
conductance peaks gradually increases as seen from both the plots of 
Fig.(\ref{fig12}). So it can be concluded that from the sharpness of 
conductance spectra, 
the relative size of mixing of spin singlet and triplet components of the gap 
parameter can be inferred.  Our result is highly in accordance with the results
of Ref. \cite{linder}. 

\section{Summary and conclusions}
In summary, in this paper we have investigated the conductance spectra
at the interface of F$|$NCSC of a F$|$S$|$F spin valve using an extended 
Blonder - Tinkham - Klapwijk (BTK) approach and the scattering matrix 
formalism. We developed the Bogoliubov de Gennes (BdG) Hamiltonian for such a  
hybrid structure introducing  the RSOC and the arbitrary orientation of the 
magnetization. We have demonstrated conductance spectra for an experimentally 
realistic parameter set, which suggest a high transparency and a moderate RSOC.
Though low value of FWM is generally preferred, however we have considered 
both low and high values of FWM for our analysis. Many conclusions can be 
drawn from our analysis. It is seen that the conductance spectra is orientation
dependent. The barrier transparency and RSOC play a very significant role. 
RSOC in general suppresses the conductance for a transparent barrier with any 
arbitrary orientation of the magnetization, however for a partially opaque 
barrier it monotonically rises with the increase of RSOC. The conductance 
becomes maximum for mid values of RSOC and then gradually fall for higher 
values of RSOC. For a strongly opaque barrier the conductance spectrum is 
found to be nearly independent of RSOC. Moreover, it is observed that
the strength of magnetization and its orientations also plays a very important 
role in the conductance spectrum. The charge conductance characteristics is 
found to be as orientation dependent. The presence unconventional 
superconductor and for different FWM, a zero bias conductance peak (ZBCP) and 
dip (ZBCD) is observed. Another important point is that for a barrier with 
high transparency, the rise of strength of magnetization enhances the 
conductance, however Andreev reflection decreases which shows a decrease 
in conductance as magnetization approaches the Fermi energy. For an 
opaque barrier the rise of strength of magnetization suppresses the charge 
conductance. It is also observed that with the rise of singlet-triplet mixing 
ratio the conductance decreases.

As a concluding remark, the results of our work indicates that a highly 
transparent spin active barrier with moderate RSOC and moderate strength of 
magnetization having an arbitrary orientation is highly suitable for a 
practical nano spin valve involving a noncentrosymmetric superconductor. We 
sincerely hope that our results shed some light on ferromagnet and 
noncentrosymmetric superconductor hybrid structures which can be utilized to 
make practical superconducting spintronic devices in near future.

\appendix
\section{Calculation of reflection and transmission coefficients}
Using BTK formalism and boundary conditions (\ref{eq17}) and (\ref{eq18}) in
the wavefunctions $\Psi_{FM}(x)$ and $\Psi_{SC}(x)$ given in Eqs.(\ref{eq12}) 
and (\ref{eq13}) respectively, we obtain $8$ linear equations connecting the 
reflection and transmission coefficients. The analytic expressions for 
reflection and the transmission coefficients can be obtained by solving the
equation of the form $x = \mathcal{A}^{-1}\mathcal{B}$, where $\mathcal{A}$ is 
a $8 \times 8$ matrix, while $\mathcal{B}$ is $8 \times 1$ matrix and $x = 
(r_e^{\uparrow},r_e^{\downarrow},r_h^{\uparrow},r_h^{\downarrow},
t_e^{\uparrow},t_e^{\downarrow},t_h^{\uparrow},t_h^{\downarrow})^T$. However,
the form of these expressions are very complex. For example, the expressions 
for the reflection coefficients can be written in the following from:
 
\begin{widetext}
\begin{eqnarray}
\label{A1}
\label{A2}
\label{A3}
\label{A4}
r_e^{\uparrow} = \frac{\omega _+^2(R_1 R_2-R_6-R_7)}{\omega _+^2P_4 (Q_7 Q_{10} \omega _-
+R_7)(O_2 O_{15} P_3 R_9-P_2 Q_1 Q_7 \omega _-)+\omega _+^2R_1 R_{14}},\\
r_e^{\downarrow}  = \frac{R_{15}+\omega _+P_4^2R_{8} 
R_{13}}{ \omega _+^2P_4 (Q_7 Q_{10} \omega _-+R_7)(O_2 O_{15} P_3 
R_9-P_2 Q_1 Q_7 \omega _-)+\omega _+^2R_1 R_{14}},\\
r_h^{\uparrow}  = \frac{R_{9}-R_{10}
-R_{16}-R_{17}}
{\omega _+^2P_4 (Q_7 Q_{10} \omega _-+R_7)(O_2 O_{15} P_3 
R_9-P_2 Q_1 Q_7 \omega _-)+\omega _+^2R_1 R_{14}},\\
r_h^{\downarrow}  = \frac{R_{12}-R_{19}+\omega _+^2P_4R_8(R_{11}-R_{18})}
{\omega _+^2P_4 (Q_7 Q_{10} \omega _-+R_7)(O_2 O_{15} P_3 
R_9-P_2 Q_1 Q_7 \omega _-)+\omega _+^2R_1 R_{14}},
\end{eqnarray}
\end{widetext}
where we define,\\
$\omega_\pm = \frac{u_{\pm}}{v_{\pm}},$\\
$O_1 = p_3-x_1$, \indent\indent\hspace{0.2mm} $O_2 = p_1-p_3$, \indent\indent $O_3 = x_5-p_3$,\\
$O_4 = p_3-x_3$, \indent\indent\hspace{0.2mm} $O_5 = p_2-x_3$, \indent\indent $O_6 = p_2-x_1$,\\
$O_7 = p_2-x_5$, \indent\indent\hspace{0.2mm} $O_8 = p_4-x_3$, \indent\indent $O_9 = p_4-x_5$,\\
$O_{10} = p_3-x_5$, \indent\indent $O_{11} = p_3+x_4$,\indent\indent $O_{12} = p_4+x_4$,\\
$O_{13} = p_2+x_4$, \indent\indent $O_{14} = p_4-x_1$, \indent\indent $O_{15} = x_5-x_3$,\\\\
$P_1 = \beta  y_1-y_2$, \indent\indent $P_2 = \beta  y_1+y_3$, \indent\indent $P_3 = \beta  y_3-y_1$,\\ 
$P_4 = \beta  y_1+y_2$, \indent\indent $P_5 = \beta  y_2+y_1$, \indent\indent $P_6 = \beta  y_1-y_3$,\\ 
$P_7 = y_1-\beta  y_2$, \indent\indent $P_8 = y_1-\beta  y_3$, \indent\indent $P_9 = \beta  y_3+y_1$,\\\\
$Q_1 = O_3 x_3 \omega _+^2+O_4 x_5+p_1 \lbrace p_3 (\omega _+^2-1)-x_5 \omega _+^2+x_3\rbrace$,\\
$Q_2 = O_{10} x_4 \omega _+^2+O_{11} x_5+p_1\lbrace p_3(\omega _+^2-1)-x_5 \omega _+^2-x_4\rbrace$,\\
$Q_3 = -(\beta ^2-1) y_2 y_1+\beta  y_1^2-\beta  y_2^2$,\\
$Q_4 = \omega _-(O_3 x_1 \omega _+^2+O_1 x_5+p_1 \lbrace p_3 (\omega _+^2-1)-x_5 \omega _+^2+x_1)\rbrace$,\\
$Q_5 = p_3 \lbrace-2 \beta  p_4 (y_1^2+y_2 y_3)+P_4 P_9 x_3+P_1 P_3 (-x_5)\rbrace
\\\indent\indent+p_4 (P_4 P_9 x_5-P_1 P_3 x_3)-2   x_3 x_5 (y_1^2+y_2 y_3)$,\\
$Q_6 = O_{10} O_{12} P_4 P_5 \omega _++O_9 O_{11} P_1 P_7 \omega _-$,\\
$Q_7 = O_4 O_9 P_1 P_8+O_8 O_{10} P_4 P_9$,\\
$Q_8 = O_1 O_9 P_1 P_2 \omega _--O_{14} O_{10} P_4 P_6 \omega _+$,\\
$Q_9 = O_8 O_{11} P_7 P_9 \omega _--O_4 O_{12} P_5 P_8 \omega _+$,\\
$Q_{10} = -O_{11} O_7 P_1P_5-O_{13} O_{10} P_4 P_5 \omega _- \omega _+$,\\\\\\
$R_1 = Q_2 Q_3 Q_7 \omega _--O_2 O_{15} P_4 P_3 Q_6$,\\
$R_2  = Q_7 \omega _- (O_6 O_{10} P_4 P_6 \omega _- \omega _+-O_1 O_7 P_1 P_2)
\\\indent\indent-Q_8(O_4 O_7 P_1 P_3-O_5 O_{10} P_4 P_9)$,\\
$R_3 = O_4 O_9 P_1 P_2 \omega _--O_8 O_{10} P_4 P_6 \omega _++P_2 Q_1 Q_7 \omega _-$,\\
$R_4 = O_4 O_9 P_1 P_2 \omega _-$,\\
$R_5 = O_2 O_{15} P_3 Q_8+P_2 Q_4 Q_7$,\\
$R_6 = Q_{10} Q_7O_2 O_{15} P_3 Q_8 \omega _--P_2 Q_4 Q_7$,\\
$R_7 = -Q_6(O_4 O_7 P_1 P_3-O_5 O_{10} P_4 P_9)$,\\
$R_8 = R_1 R_2 \omega _+^2+R_6+R_7$,\\
$R_9 = R_4-O_8 O_{10} P_4 P_6 \omega _+$,\\
$R_{10} = \frac{O_1 O_9 P_1 P_2 \omega _--O_{14} O_{10} P_4 P_6 \omega _+}{Q_7 \omega _- \omega _+}$,\\
$R_{11} = \frac{O_4 P_8 R_9}{O_{10} P_4 Q_7 \omega _- \omega _+}-\frac{O_4 P_2}{O_{10} P_4 \omega _+}$,\\
$R_{12} = \frac{O_1 P_2}{O_{10} P_4 \omega _+}-\frac{O_4 P_8 Q_8}{O_{10} P_4 Q_7 \omega _- \omega _+}$,\\
$R_{13} = -O_2 O_{15} P_3 R_3$,\\
$R_{14}=Q_7 \omega _-(O_5 O_{10} P_4 P_6 \omega _- \omega _+-O_4 O_7 P_1 P_2)
\\ \indent\indent-R_9(O_4 O_7 P_1 P_3-O_5 O_{10} P_4 P_9)$,\\
$R_{15} = -\frac{P_4 R_5}{R_1}$,\\
$R_{16} = \frac{P_4^2 Q_6 R_{13}R_8 \omega _+}{R_1Q_7 \omega _- }$,\\
$R_{17} =  \frac{P_4 Q_6 R_5}{Q_7 R_1 \omega _- \omega _+}$,\\
$R_{18} =  \frac{P_4 Q_9 R_{13}}{Q_5 R_1 \omega _- \omega _+}$,\\
$R_{19} =  \frac{P_4 Q_9 R_5}{Q_5 R_1 \omega _- \omega _+}$,\\\\
$\beta = e^{-i\phi}$, \indent\indent\hspace{0.2mm}\\ 
$x_1 = k^+_F\cos\theta_F - 2iU_{int}$,\\
$x_2 = k^-_F\cos\theta_F - 2iU_{int}$, \\ 
$x_3 = k^+_FS_1 - 2iU_{int}$,\\
$x_4 = k^-_FS_2 + 2iU_{int}$, \\
$x_5 = k^-_FS_2 - 2iU_{int}$,\\
$p_1 = \lambda\cos\theta_e$,\indent\indent\hspace{0.2mm} $p_2 = \lambda \cos\theta_e$,\\
$p_3 = \lambda\cos\theta_h$,\indent\indent\hspace{0.2mm} $p_4 = \lambda\cos\theta_h$,\\
$y_1 = \cos\theta_m$, \hspace{0.2mm} $y_2 = \sin\theta_m e^{i \chi_m}$,  \hspace{0.2mm} $y_3 = \sin\theta_m e^{-i \chi_m}$.\\

\end{document}